\documentclass[reprint,]{revtex4-2}
\usepackage{color,amsmath,amssymb,commath,graphicx}

\begin{document}

\title{Mechanical response in elastic fluid flow networks}
\author{Sean Fancher}
\email{sfancher@sas.upenn.edu}
\affiliation{Department of Physics and Astronomy, University of Pennsylvania, Philadelphia, PA 19104, USA}
\author{Eleni Katifori}
\affiliation{Department of Physics and Astronomy, University of Pennsylvania, Philadelphia, PA 19104, USA}

\begin{abstract}

The dynamics of flow within a material transport network is dependent upon the dynamics of its power source. Responding to a change of these dynamics is critical for the fitness of living flow networks, e.g. the animal vasculature, which are subject to frequent and sudden shifts when the pump (the heart) transitions between different steady states. The combination of flow resistance, fluid inertia, and elasticity of the vessel walls causes the flow and pressure of the fluid throughout the network to respond to these transitions and adapt to the new power source operating profiles over a nonzero time scale. We find that this response time can exist in one of two possible regimes; one dominated by the decay rate of travelling wavefronts and independent of system size, and one dominated by the diffusive nature of the fluid mechanical energy over large length scales. These regimes are shown to exist for both single vessels and hierarchically structured networks with systems smaller than a critical size in the former and larger systems in the later. Applying biologically relevant parameters to the model suggests that animal vascular networks may have evolved to occupy a state within the minimal response time regime but close to this critical system size.

\end{abstract}

\maketitle

\section{Introduction}
\label{sec:intro}

A wide variety of both natural and artificial systems can be described as a material transport network. In the most general sense, such networks are defined by their topological structure, internal forces and flow dynamics, and a set of boundary conditions set by the power source. In cases such as when the transported fluid is incompressible and the tubes through which it flows are rigid, the internal dynamics of the flow happen on a negligibly short time scale and the system can be completely defined by its topology and boundary conditions. This is exemplified in water distribution systems, where the pressure at the distribution nodes responds effectively immediately relative to gradual changes in water level at the water towers over the course of a day. However, when the internal dynamics themselves occur over time scales similar to or greater than those of the boundary conditions, the system can gradually respond to the externally imposed changes, and the manner in which it does so can become a crucial system design aspect that can be optimized.

Many systems can be forced into this regime by simply implementing boundary conditions with sufficiently rapid dynamics. This is seen in commercial water distribution networks via the hydraulic shock or ``water hammer'' phenomenon, which can cause substantial damage to the system \cite{ghidaoui2005review}. Conversely, it is possible to use less extreme boundary conditions and slow the internal dynamics by allowing the channels through which the fluid flows to be compliant and capable of storing excess volume. This is seen in animal vasculature in which blood vessels can expand to accommodate increased blood volume \cite{sherwin2003one,alastruey2012physical}. Indeed, many previous studies have investigated the effects of vessel compliance on flow and pressure waveforms throughout the body \cite{holenstein1988reverse,sherwin2003one,bui2009dynamics,alastruey2012physical,pan2014one,perdikaris2015effective,flores2016novel,yigit2016non,bauerle2020living}, but these have typically been restricted to subsections of the whole vasculature and/or flows under steady state flow input profiles with periodic boundary conditions in time.

Contrary to these modelling practices, naturally occurring flow networks such as the animal vasculature are frequently disrupted away from a given dynamical steady state by sudden changes in boundary conditions. Not only can the animal's heart rate shift to create increased or decreased blood flow as needed, but the blood vessels themselves can become dilated, constricted, or damaged. In each of these instances, the flow and pressure throughout the network, and thus the rate at which nutrients are delivered to the body's cells, is affected. While many animals, including humans, are capable of locally controlling and redistributing blood flow via mechanisms such as vascular smooth muscle externally applying pressure \cite{olufsen2005blood,pries2015coronary}, the total flow rate of blood throughout the entire body is typically managed by the dynamics of the heart itself. As such, the mechanical limits on how quickly the flow at any arbitrary point within the body can adapt to changes in heart rate can set bounds on the organism's ability to respond to sudden external stimuli.

Here, we investigate the response of flow within a material transport network comprised of compliant vessels to changes in boundary conditions. In Sec.~\ref{sec:singlevessel} we obtain a set of dynamic equations for the pressure and flow within a single vessel by linearizing the Navier-Stokes equations for flow within an elastic, cylindrical tube \cite{barnard1966theory,sherwin2003one,alastruey2012physical,cousins2013new}. The resulting equations are a special form of the telegrapher's equations \cite{masoliver1994telegrapher}, from which we construct networks with well defined connectivity laws between vessels. We find that there exists a minimum possible time scale over which both single vessels and whole networks are capable of responding to a sudden change in boundary conditions that is dictated by the decay rate of wavefront amplitudes. There also exists a critical size above which the vessel or network will respond more slowly than this minimum due to the mechanical energy propagating in a diffusive manner over large length scales. For single vessels we are able to solve the dynamic equations analytically and directly calculate how these response behaviors depend on the vessel parameters. For whole networks we use numerical integration defined in Sec.~\ref{sec:networks} to show the same results hold given network averaged parameters analogous to those of the single vessel. Finally, we obtain a generalized method for approximating the time scale over which the flow and pressure will adapt to a given set of changes in the boundary conditions of a network. Our work highlights the importance of the response time, the time for the network to adapt to the new pump flow conditions, as an important design consideration for networks composed of elastic vessels.

\section{Results}

\subsection{Single Vessel Mechanics}
\label{sec:singlevessel}

\begin{figure}[t]
    \centering
    \includegraphics[width=\columnwidth]{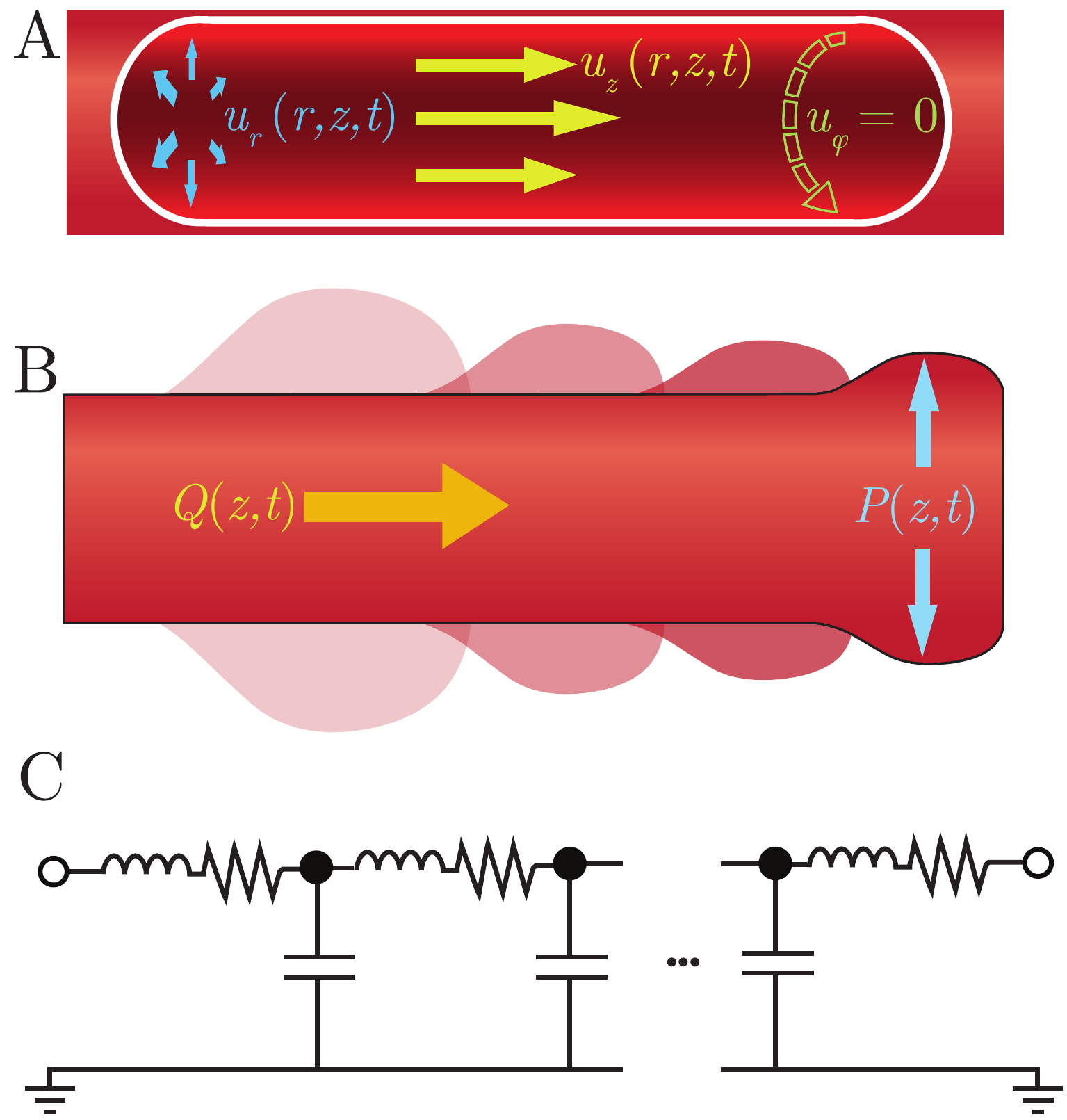}
    \caption{Diagrams of 3D and 1D flow and pressure mechanics. A) Fluid flowing through a compliant cylindrical vessel with axial and radial velocities, $u_{z}$ and $u_{r}$, cause changes in the radius and cross sectional area of the vessel in both space and time. Rotational symmetry enforces $u_{\phi}=0$. B) The radial component of the flow is integrated out so as to write the dynamics purely in terms of the volumetric flow, $Q$, and fluid pressure, $P$, dependent only on time and the axial dimension. Current and pressure pulses can travel through the vessel with dynamics dictated by Eqs. \ref{Qsym} and \ref{Psym}, resulting in exponentially decaying pulse amplitudes. C) The flow dynamics resulting from Eqs. \ref{Qsym} and \ref{Psym} can also be interpreted as the continuum limit of a series of inductors and resistors connected in parallel to a ground through capacitors.}
    \label{diagrams}
\end{figure}

We begin by considering an incompressible fluid with density $\rho$ and viscosity $\mu$ flowing through a cylindrical vessel (Fig. \ref{diagrams}A). We assume the system is rotationally symmetric so that the flow rate and fluid pressure depend only on the axial and radial positions $z$ and $r$. Denoting the axial and radial fluid velocities as $u_{z}(z,r,t)$ and $u_{r}(z,r,t)$ respectively and the fluid pressure as $p(z,r,t)$, the incompressibility condition and Navier-Stokes equation are \cite{barnard1966theory,sherwin2003one,alastruey2012physical,cousins2013new}

\begin{equation}
    \vec{\nabla}\cdot\vec{u} = \frac{\partial u_{z}}{\partial z}+\frac{1}{r}\frac{\partial}{\partial r}\left(ru_{r}\right) = 0,
    \label{incom}
\end{equation}

\begin{equation}
    \vec{\nabla}p+\rho\frac{\partial\vec{u}}{\partial t}+\rho\left(\vec{u}\cdot\vec{\nabla}\right)\vec{u}-\mu\nabla^{2}\vec{u} = 0.
    \label{NSeq}
\end{equation}

We now reexpress Eqs. \ref{incom} and \ref{NSeq} in terms of the total volumetric flow rate $Q(z,t)=\int dA\; u_{z}(z,r,t)$ and area averaged pressure $P(z,t)=A(z,t)^{-1}\int dA\; p(z,r,t)$, where the integration is performed over the cross sectional area, $A(z,t)$, of the vessel at axial position $z$ and time $t$ (Fig. \ref{diagrams}B). We first integrate the incompressibility condition (Eq. \ref{incom}) and average the axial component of the Navier-Stokes equation (Eq. \ref{NSeq}) over the vessel cross section to produce equations for $\partial Q/\partial z$ and $\partial P/\partial z$. By equating the radial fluid velocity at the vessel wall to the expansion rate of the wall, the second term in Eq. \ref{incom} can be shown to simply become $\partial A/\partial t$ once this integration is performed. By restricting our system to the regime in which the Womersely number is small, we can use the laminar flow solution for the fluid velocity. This allows the radial term in $-\nabla^{2}u_{z}$ to be reexpressed as $8\pi Q/A^{2}$ once it is area averaged. The axial term is negligible whenever the wavelength of any pulses travelling through the fluid is significantly larger than the vessel radius, which is another restriction we impose. The nonlinear term, $\rho(\vec{u}\cdot\vec{\nabla})\vec{u}$, can be broken up into two distinct pieces that are both made to be negligible; one by the assumption that wavelengths are longer than the vessel radius while the other by the assumption that the fluid velocity is much slower than propagation velocity of such pulses. The culmination of these manipulations and approximations is given in Appendix \ref{LNSE} and transforms Eqs. \ref{incom} and \ref{NSeq} into

\begin{equation}
    \frac{\partial Q}{\partial z}+\frac{\partial A}{\partial P}\frac{\partial P}{\partial t} = 0,
    \label{Qeq}
\end{equation}

\begin{equation}
    \frac{\partial P}{\partial z}+\frac{\rho}{A}\frac{\partial Q}{\partial t}+\frac{8\pi\mu}{A^{2}}Q = 0.
    \label{Peq}
\end{equation}

Eq. \ref{Qeq} can be simplified by assuming that the vessel cross sectional area scales linearly with the fluid pressure as $A(z,t)=A_{0}+cP(z,t)$, where $c$ is the compliance of the vessel. Additionally, we make the assumption that the vessel cross section, $A(z,t)$, does not significantly change ($A_{0}\gg cP(z,t)$) so as to allow the factors of $A$ in Eq. \ref{Peq} to be sufficiently approximated by the constant $A_{0}$. We can now define the fluid inertia and the flow resistance per unit length as $\ell=\rho/A_{0}$ and $r=8\pi\mu/A_{0}^{2}$ respectively. These three parameters, $c$, $\ell$, and $r$, thus characterize the system and allow us to define three distinct derived parameters: the characteristic length scale $\lambda=2(\sqrt{\ell/c})/r$, the characteristic time scale $\tau=2\ell/r$, and the characteristic admittance $\alpha=\sqrt{c/\ell}$. Reformulating Eqs. \ref{Qeq} and \ref{Peq} to be expressed in terms of these characteristic parameters produces the more symmetric looking versions

\begin{equation}
    \lambda\frac{\partial Q}{\partial z}+\alpha\tau\frac{\partial P}{\partial t} = 0,
    \label{Qsym}
\end{equation}

\begin{equation}
    \alpha\lambda\frac{\partial P}{\partial z}+\tau\frac{\partial Q}{\partial t}+2Q = 0.
    \label{Psym}
\end{equation}

Eqs. \ref{Qsym} and \ref{Psym} represent a simple form of the telegrapher's equations with spatially and temporally independent parameters \cite{masoliver1994telegrapher}. The derivative terms create traveling waves of current and pressure while the existence of the resistive term $2Q$ in Eq. \ref{Psym} causes the waves to exponentially decay, as depicted in Fig. \ref{diagrams}B. These equations can also be derived via an analogous transmission line circuit with no shunt resistor, as depicted in Fig. \ref{diagrams}C. In total, the necessary assumptions required to obtain Eqs. \ref{Qsym} and \ref{Psym} from Eqs. \ref{incom} and \ref{NSeq} are that the system is rotationally symmetric, the Womersley number is sufficiently small that the fluid velocity profile is approximately that of Poiseuille flow, the flow velocity is sufficiently small compared to the velocity of current and/or pressure pulses, the wavelength and/or exponential length scale of the flow is sufficiently large compared to vessel radius, and changes to the vessel cross sectional area are small and approximately linear with changes in pressure. We expand on the description of each of these assumptions in Appendix \ref{LNSE} and comment on their validity as they pertain to biological contexts in Sec. \ref{sec:disc}.

To obtain a set of solutions to Eqs. \ref{Qsym} and \ref{Psym} we first consider the function $W(z,t)$ defined such that $Q(z,t)=\tau\partial_{t}W(z,t)$. Inserting this into Eq. \ref{Qsym} then dictates that the quantity $\alpha P(z,t)+\lambda\partial_{z}W(z,t)$ must vanish when it is differentiated with respect to $t$ and thus be a function only of $z$. However, since $W(z,t)$ can still satisfy its defining equation $Q(z,t)=\tau\partial_{t}W(z,t)$ when any time independent function is added to it, we are free to choose a $W(z,t)$ such that $\alpha P(z,t)+\lambda\partial_{z}W(z,t)=0$. These two conditions uniquely specify $W(z,t)$ up to an additive constant and allow Eq. \ref{Psym} to be written as

\begin{equation}
	-\lambda^{2}\frac{\partial^{2}W}{\partial z^{2}}+\tau^{2}\frac{\partial^{2}W}{\partial t^{2}}+2\tau\frac{\partial W}{\partial t} = 0.
	\label{Wsym}
\end{equation}

\noindent By differentiating Eq. \ref{Wsym} with respect to $t$ or $z$, the same equation for $Q$ and $P$ respectively can also be obtained, thus implying that any solution for $W$ is also a possible solution for $Q$ or $P$ under different boundary conditions. Here, we choose to work with $W$ as obtaining $Q$ and $P$ from it is relatively simple whereas obtaining $P$ from a solution to $Q$ or vice versa can be notably more complex.

One method of solving Eq. \ref{Wsym} is to factor a $\text{exp}(-t/\tau)$ out of $W(z,t)$ then reexpress the remaining function in terms of the new independent variables $q(z,t)=\sqrt{(t/\tau)^{2}-(z/\lambda)^{2}}$ and $s(z,t)=\sqrt{(t/\tau-z/\lambda)/(t/\tau+z/\lambda)}$. Assuming separation of variables holds in $q-s$ space gives the dimensionless solution set (see supplemental material)

\begin{equation}
	W_{n}\left(z,t\right) = e^{-\frac{t}{\tau}}s^{n}\left(z,t\right)I_{n}\left(q\left(z,t\right)\right),
	\label{Wndef}
\end{equation}

\noindent where $I_{n}(x)$ is the $n$th modified Bessel function of the first kind. Replacing $I_{n}(x)$ with $K_{n}(x)$, the $n$th modified Bessel function of the second kind, is also a valid solution, but here we will work exclusively with those solutions given by Eq. \ref{Wndef}.

We now consider a semi-infinite vessel that exists on the interval $z\in [ 0,\infty )$. We first impose the initial conditions $Q(z,t<0)=0$ and $P(z,t<0)=0$ along with the current boundary condition $Q(z=0,t)=\hat{Q}\delta(t/\tau)$, or equivalently $W(z=0,t)=\hat{Q}\Theta(t/\tau)$, where $\delta(x)$ is the Dirac $\delta$-function and $\Theta(x)$ is the Heaviside step function. After multiplying by $\text{exp}(-t/\tau)\text{exp}(t/\tau)$, the factor of $\text{exp}(t/\tau)$ can be expanded using the generating function for the modified Bessel functions. This process produces the solution $W(z,t)=\hat{Q}\Theta(t/\tau-z/\lambda)(W_{0}(z,t)+2\sum_{n=1}^{\infty}W_{n}(z,t))$, from which $Q(z,t)$ and $P(z,t)$ can be derived. This solution can then be utilized as a kernel function for any arbitrary current boundary condition $Q(z=0,t)=H(t)$ to produce the current solution

\begin{align}
	&Q\left(z,t\right) = H\left(t-\frac{z\tau}{\lambda}\right)e^{-\frac{z}{\lambda}} \nonumber\\
	&+\int_{\frac{z\tau}{\lambda}}^{\infty}\frac{dt'}{\tau}\> H\left(t-t'\right)\frac{1}{2}\left(W_{-1}\left(z,t'\right)-W_{1}\left(z,t'\right)\right) \nonumber\\
	&= H\left(t-\frac{z\tau}{\lambda}\right)e^{-\frac{z}{\lambda}}+\int_{\frac{z\tau}{\lambda}}^{\infty}dt'\> H\left(t-t'\right)\frac{zI_{1}\left(q\left(z,t'\right)\right)}{\tau\lambda e^{\frac{t'}{\tau}}q\left(z,t'\right)}.
	\label{Qgensol}
\end{align}

The two terms in Eq. \ref{Qgensol} have very distinct interpretations. The first term represents the current pulse generated by the boundary condition travelling with a finite velocity of $\lambda/\tau$. This effect is due to the hyperbolic nature of the first two terms of Eq. \ref{Wsym} restricting the propagation speed of disturbances in $W$, and in turn $Q$, to exactly this value. Additionally, the inclusion of the dissipative third term in Eq. \ref{Wsym} causes the resulting current pulse to exponentially decay with distance travelled as it loses energy to friction as well as wave dispersion. It is the second term of Eq. \ref{Qgensol} that shows precisely how this dispersion of the current pulse occurs over the length of the vessel occurs. In the long time limit ($t'/\tau\gg z/\lambda$), this spreading can be seen to be approximately diffusive as $I_{1}(x)$ can be replaced with its large argument limit $\text{exp}(x)/\sqrt{2\pi x}$ and $q(z,t')$ can be expanded to lowest order in $z$:

\begin{align}
	&\frac{zI_{1}\left(q\left(z,t'\right)\right)}{\tau\lambda e^{\frac{t'}{\tau}}q\left(z,t'\right)} \approx \frac{ze^{q\left(z,t'\right)-\frac{t'}{\tau}}}{\tau\lambda\sqrt{2\pi}\left(q\left(z,t'\right)\right)^{3/2}} \nonumber\\
	&\approx 2\frac{2z}{4t'}\frac{1}{\sqrt{4\pi\left(\lambda^{2}/2\tau\right)t'}}e^{-\frac{z^{2}}{4\left(\lambda^{2}/2\tau\right)t'}} \nonumber\\
	&= -2\frac{\lambda^{2}}{2\tau}\frac{\partial}{\partial z}\left(\frac{1}{\sqrt{4\pi\left(\lambda^{2}/2\tau\right)t'}}e^{-\frac{z^{2}}{4\left(\lambda^{2}/2\tau\right)t'}}\right).
	\label{diff_kernel}
\end{align}

\noindent By defining $D=\lambda^{2}/2\tau$, the final form of Eq. \ref{diff_kernel} is of the form $-2D\partial_{z}\text{exp}(-z^{2}/4Dt)/\sqrt{4\pi Dt}$, which is the expression for the flow of diffusive material over a one dimensional semi-infinite domain. Thus, the second term of Eq. \ref{Qgensol} can be interpreted as approximately representing a diffusive spreading of the boundary condition over the vessel after a sufficiently long time. This understanding is reinforced by the fact that a current pulse from far in the past will have substantially dispersed over the vessel and its contribution to the current will be small and changing very slowly with time. This allows the second term of Eq. \ref{Wsym} to be neglected after such a long time, thus producing the diffusion equation with precisely the same value of the diffusion constant, $D=\lambda^{2}/2\tau$.

\begin{figure}[t]
    \centering
    \includegraphics[width=\columnwidth]{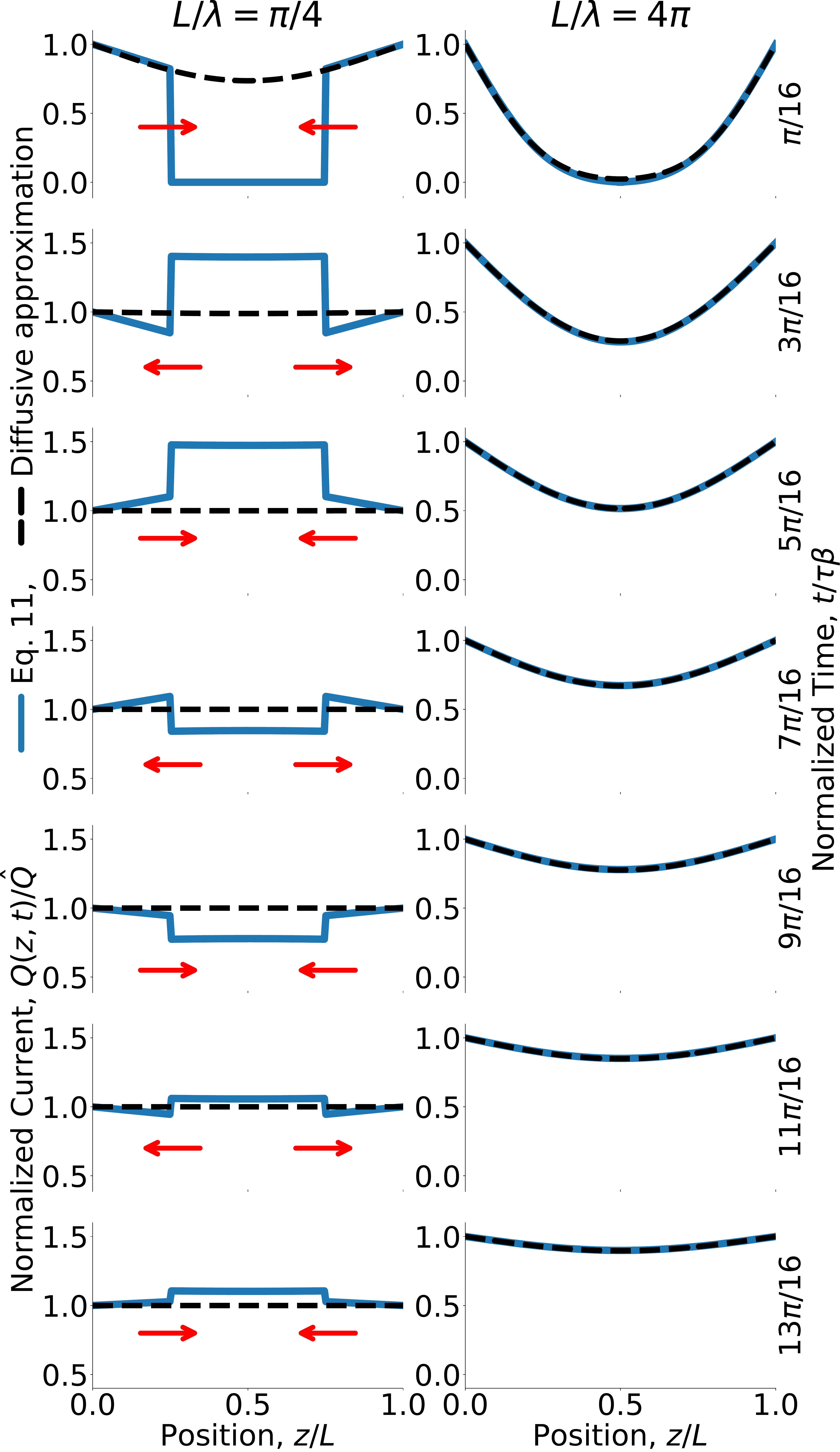}
    \caption{Normalized current, $Q(z,t)/\hat{Q}$, at various points in time for vessels of size $L/\lambda=\pi/4$ and $L/\lambda=4\pi$ with step function boundary conditions $Q(0,t)=Q(L,t)=\hat{Q}\Theta(t/\tau)$. In each case, the blue curves represent the solution given by Eq. \ref{Qsymsol} while the dashed black curves represent the solution to the diffusion equation with $D=\lambda^{2}/2\tau$ and the same step function boundary conditions. For the short vessel (left), red arrows indicate the direction of travel for the decaying wavefronts created by the boundary conditions. For the long vessel (right), equivalent values of normalized time, $t/\tau\beta$, represent longer actual times as $\beta>1$ in this regime by Eq. \ref{betadef}. The depicted current dynamics are also shown in supplemental videos.}
    \label{Qframes}
\end{figure}

With the solution for the semi-infinite vessel, we can obtain a solution for a finite vessel of length $L$ with arbitrary boundary conditions at either end via the method of images. Here, we will specifically focus on the case in which $Q(z=0,t)=Q(z=L,t)=\hat{Q}\text{exp}(i\omega t)\Theta(t/\tau)$. Given these boundary conditions, we can express the modified Bessel function in Eq. \ref{Qgensol} as an integral of the generating function around the unit circle of the complex plane. This allows the summation over the images to be performed and the residue theorem to be applied to the resultant infinite series of poles, ultimately leading to the solution (see supplemental material)

\begin{align}
	&Q\left(z,t\right) = \hat{Q}\left(e^{i\omega t}\frac{\sinh\left(\frac{z}{\lambda}k\left(\omega\tau\right)\right)+\sinh\left(\frac{L-z}{\lambda}k\left(\omega\tau\right)\right)}{\sinh\left(\frac{L}{\lambda}k\left(\omega\tau\right)\right)}\right. \nonumber\\
	&-\left.e^{-\frac{t}{\tau}}\sum_{m\in\mathbb{O}^{+}}\frac{4\pi m\sin\left(\pi m\frac{z}{L}\right)\Omega_{m}\left(\frac{t}{\tau},\frac{L}{\lambda},\omega\tau\right)}{\pi^{2}m^{2}+\left(\frac{L}{\lambda}k\left(\omega\tau\right)\right)^{2}}\right),
	\label{Qsymsol}
\end{align}

\begin{align}
    &\Omega_{m}\left(\frac{t}{\tau},\frac{L}{\lambda},\omega\tau\right) \nonumber\\
    &= \begin{cases}
    \frac{1+i\omega\tau}{\sqrt{1-\left(\frac{\pi m}{L/\lambda}\right)^{2}}}\sinh\left(\frac{t}{\tau}\sqrt{1-\left(\frac{\pi m}{L/\lambda}\right)^{2}}\right) & \\
    \quad +\cosh\left(\frac{t}{\tau}\sqrt{1-\left(\frac{\pi m}{L/\lambda}\right)^{2}}\right) & \frac{L/\lambda}{\pi m} > 1 \\
    1+\frac{t}{\tau}\left(1+i\omega\tau\right) & \frac{L/\lambda}{\pi m}=1 \\
    \frac{1+i\omega\tau}{\sqrt{\left(\frac{\pi m}{L/\lambda}\right)^{2}-1}}\sin\left(\frac{t}{\tau}\sqrt{\left(\frac{\pi m}{L/\lambda}\right)^{2}-1}\right) & \\
    \quad +\cos\left(\frac{t}{\tau}\sqrt{\left(\frac{\pi m}{L/\lambda}\right)^{2}-1}\right) & \frac{L/\lambda}{\pi m} < 1 \end{cases},
    \label{Omegadef}
\end{align}

\noindent where $\mathbb{O}^{+}$ is the set of all positive odd integers and $k\left(\omega\tau\right)=\sqrt{i\omega\tau(2+i\omega\tau)}$ with the principle root being taken. Eq. \ref{Qsymsol} represents a valid solution for all times $t/\tau\ge\text{min}(z,L-z)/\lambda$ with $Q(z,t)=0$ for all other times. Fig. \ref{Qframes} depicts this solution at various times for vessels of length $L/\lambda=\pi/4$ and $L/\lambda=4\pi$.

\begin{figure}[t]
    \centering
    \includegraphics[width=\columnwidth]{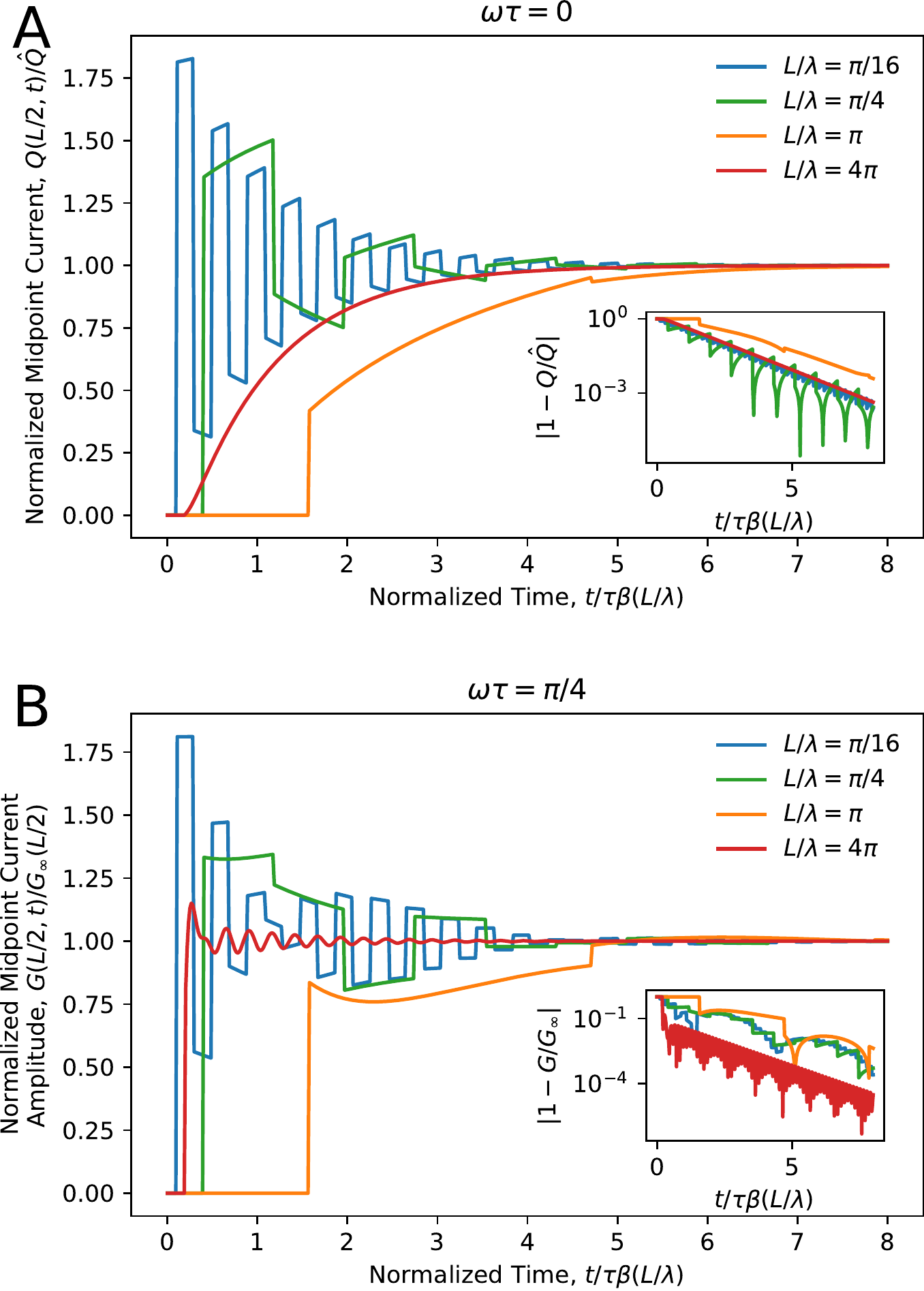}
    \caption{Current response of single vessel for step function and pulsatile step function boundary conditions. A) When symmetric, step function boundary conditions are applied to the vessel, the current at the midpoint is seen to undergo decaying oscillations if the vessel is short or simply slowly climb towards the final steady state value if the vessel is long. Inset plot shows the magnitude of the difference between the current and its final steady state value. B) For step function pulsatile boundary conditions, the current amplitude is seen to similarly undergo decaying oscillations that become smaller yet last longer as the vessel becomes longer. Inset plot similarly shows the magnitude of the difference from the final steady state value. In both cases, the inset plots show that the system exponentially approaches its long time steady state over a time scale defined by $\tau\beta(L/\lambda)$ (Eq. \ref{betadef}).}
    \label{SVresp}
\end{figure}

As with Eq. \ref{Qgensol}, Eq. \ref{Qsymsol} can be broken into two distinct terms each with their own interpretation. The first term, shown in the upper line of Eq. \ref{Qsymsol}, is the steady state term as it simply oscillates with time and thus represents the long time steady state solution of the system. The second term, given by the summation in the second line of Eq. \ref{Qsymsol}, is the response term as it exponentially decays with time and thus measures how the system adapts to the shift in boundary conditions. Importantly, the $z$ dependence of each term in this summation is relegated solely to the factor of $\sin(\pi mz/L)$. We can thus interpret each of these as measuring the contribution to the response term given by a standing wave of wavelength $2L/m$. In fact, since $Q$ must also obey Eq. \ref{Wsym} we can use this form to replace the operator $-\lambda^{2}\partial_{z}^{2}$ with multiplication by the constant $(\pi m\lambda/L)^{2}$. This transforms Eq. \ref{Wsym} into the equation for a simple damped harmonic oscillator with a natural frequency of $\pi m\lambda/(L\tau)$ and a damping ratio of $L/(\pi m\lambda)$, both of which depend on the wavelength of the standing wave they describe but not on the driving frequency, $\omega$.

Given this interpretation, we see that when the wavelength is short ($L/(\pi m\lambda)<1$) the system is underdamped and undergoes decaying oscillations. The sinusoidal nature of $\Omega_{m}$ in this regime produces these oscillations while the global factor of $\text{exp}(-t/\tau)$ in the response term of Eq. \ref{Qsymsol} accounts for the decay of the amplitude. All terms for which $m>L/(\pi\lambda)$ will behave in this underdamped manner, which includes all possible values of $m$ when the vessel is short ($L/\lambda<\pi$). If the vessel is long enough to allow for long wavelengths ($L/(\pi m\lambda)>1$) then the system is overdamped and simply decays exponentially. The hyperbolic nature of $\Omega_{m}$ in this regime accounts for the two distinct decay timescales, but after a sufficient amount of time terms that decay over the short timescale will be negligible and only those of the form $\text{exp}(-(t/\tau)(1-\sqrt{1-(\pi m\lambda/L)^{2}}))$ will remain. Only finitely many terms for which $m<L/(\pi\lambda)$ will behave in this way and only when the vessel is sufficiently long ($L/\lambda>\pi$). Finally, in the critical case ($L/(\pi m\lambda)=1$) the wave decays in a critical manner, thus causing $\Omega_{m}$ to grow linearly in time. When the vessel is precisely at the critical size $L/\lambda=\pi$, the $m=1$ term is dominant and the response term also decays critically. The culmination of these effects dictates that the response term in Eq. \ref{Qsymsol} decays approximately exponentially as $\text{exp}(-t/\tau\beta(L/\lambda))$, where the response scaling function, $\beta(L/\lambda)$, gives the time scale over which the system responds to changes in boundary conditions in units of $\tau$ and is defined as

\begin{equation}
	\beta\left(\frac{L}{\lambda}\right) = \begin{cases}
	1 & \frac{L}{\lambda}\le\pi \\
	\left(1-\sqrt{1-\left(\frac{\pi\lambda}{L}\right)^{2}}\right)^{-1} & \frac{L}{\lambda}>\pi \end{cases}.
	\label{betadef}
\end{equation}

Fig. \ref{SVresp}A shows the $\omega=0$ case of Eq. \ref{Qsymsol} at the location $z=L/2$, normalized by $\hat{Q}$. From it we see that when $L/\lambda$ is less than the critical value of $\pi$, only underdamped modes exist and the midpoint current undergoes decaying oscillations around its long time steady state value. The exact timing of these spikes in current can be understood intuitively as a consequence of the first term of Eq. \ref{Qgensol}. This term represents a wavefront travelling with velocity $\lambda/\tau$ and decaying in amplitude as $\text{exp}(-z/\lambda)$. As these wavefronts reach the center of the vessel from either side they cause the current to spike upward. Reflections off either end of the vessel then induce a change of sign and force the current to spike downwards once these reflected wavefronts return. This process, depicted in the $L/\lambda=\pi/4$ case of Fig. 2, repeats over and over with each successive reflection being decayed more and more, thus producing the pattern seen in Fig. \ref{SVresp}A. The fact that these reflecting wavefronts dominate the reponse of the system and decay at a constant rate independent of $L$ in turn causes the response scaling function, $\beta(L/\lambda)$, to take on a constant, also $L$-independent value of $1$.

Conversely, when $L/\lambda>\pi$, the vessel is long enough for overdamped modes to exist and the midpoint current simply decays exponentially towards its long time steady state value. In this regime, $\beta(L/\lambda)$ can be well approximated by the simple form $2(L/\pi\lambda)^{2}$, thus causing the time to approach steady state to increase quadratically with vessel size. This is exactly what one would expect from a diffusive system and can be seen as a consequence of the second term of Eq. \ref{Qgensol} and its diffusive approximation explored in Eq. \ref{diff_kernel}. The $L/\lambda=4\pi$ case of Fig. \ref{Qframes} reinforces this interpretation by displaying the excellent agreement between the solution to Eq. \ref{Qsymsol} and this diffusive approximation. Finally, as $L/\lambda$ approaches the critical value of $\pi$, the critically damped mode becomes extant and causes the response term to take on a slightly larger value from the linear term seen in $\Omega_{m}$. This effect is seen clearly in the inset of Fig. \ref{SVresp}A, which plots the response term itself.

When $\omega$ is nonzero, the midpoint current approaches a time dependent steady state rather than a static value. To study this case, we take the real part of Eq. \ref{Qsymsol}, which is equivalent to imposing the purely real boundary conditions $Q(z=0,t)=Q(z=L,t)=\hat{Q}\cos(\omega t)\Theta(t/\tau)$. From the steady state term of Eq. \ref{Qsymsol}, we see that at any point along the length of the vessel, the current will eventually approach a sinusoidally oscillating state. Given this, we can express the full time dependent current in the form $Q(z,t)=J(z,t)\cos(\omega t+\phi(z,t))$ for some amplitude $J(z,t)$ and phase $\phi(z,t)$ that are both dependent on space and time but asymptotically approach constant values as $t$ increases. To extract how the amplitude in particular evolves over time, we consider the function $G(z,t)=\sqrt{(Q(z,t))^{2}+(\omega^{-1}\partial_{t}Q(z,t))^{2}}$. In the regime where $\omega^{-1}\partial_{t}J(z,t)\ll J(z,t)$ and $\omega^{-1}\partial_{t}\phi (z,t)\ll 1$, which is guaranteed to happen at sufficiently long times, $G(z,t)$ approaches $J(z,t)$.

Fig. \ref{SVresp}B shows the $\omega\tau=1/4$ case of $G(z,t)$ at the midpoint $z=L/2$ and normalized by its long time limit, $G_{\infty}(z)=\lim_{t\to\infty}G(z,t)$. When $L/\lambda\le\pi$ we see qualitatively similar behaviors to the $\omega=0$ case with shorter vessels undergoing decaying oscillations of increasingly longer period as $L/\lambda$ increases. When $L/\lambda>\pi$, the response term of Eq. \ref{Qsymsol} has a noticably smaller magnitude due to the $(Lk(\omega\tau)/\lambda)^{2}$ term in the denominator dominating over $\pi^{2}m^{2}$ for small $m$, which in turn causes $G(z,t)$ to have much smaller deviations from $G_{\infty}(z)$. The inset of Fig. \ref{SVresp}B shows $G(L/2,t)$ exponentially approaching $G_{\infty}(L/2)$ over a time scale governed by the response scaling function, $\beta(L/\lambda)$. This shows that $\beta(L/\lambda)$ dictates the response time of not only shifts in the $\omega=0$ component of the current, but also the nonzero frequency components.

\begin{figure}[t]
    \centering
    \includegraphics[width=\columnwidth]{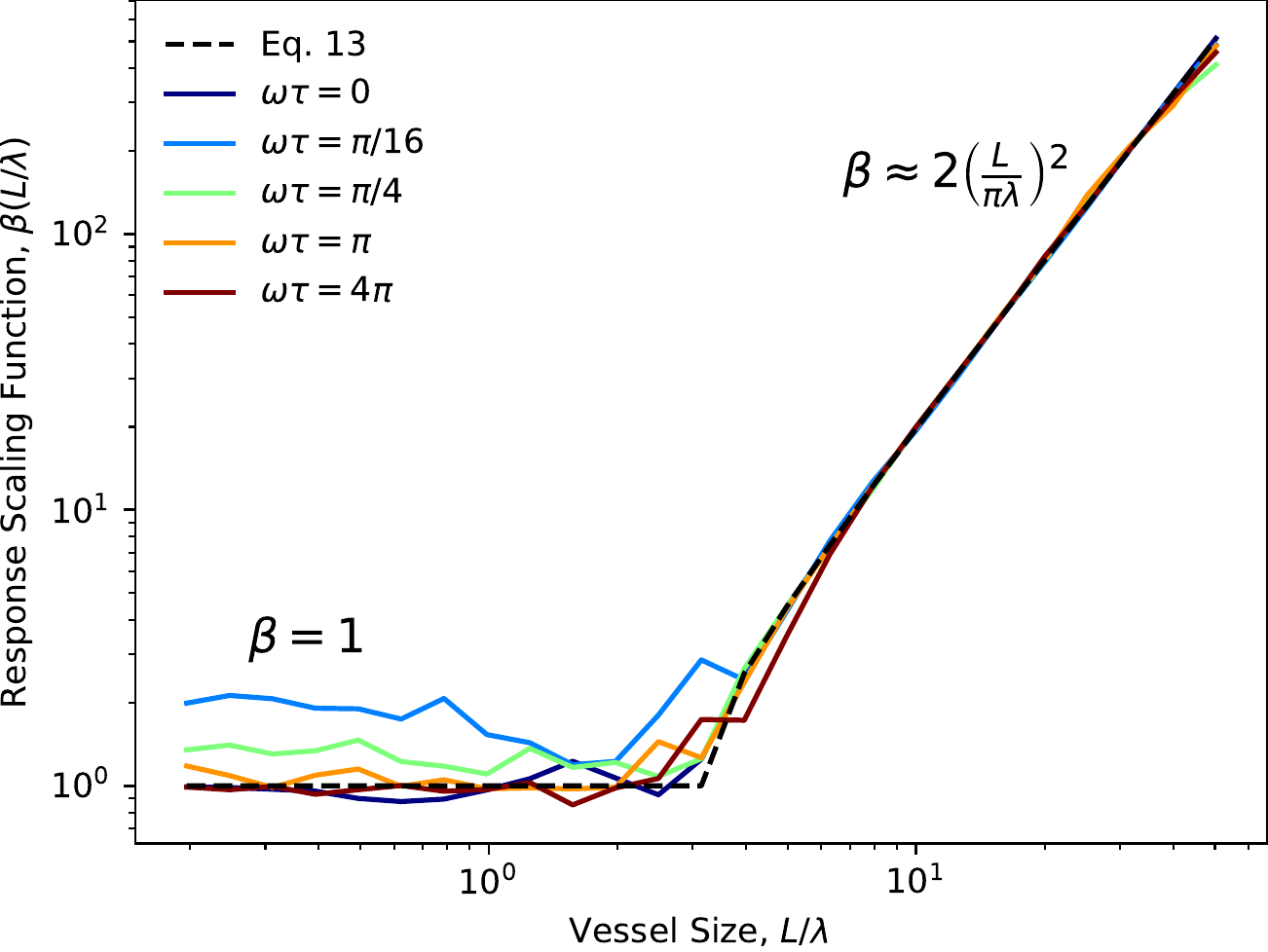}
    \caption{By performing a linear fit to the data presented in the insets of Fig. \ref{SVresp}, the numerically measured response scaling function can be directly compared to the predicted form given by Eq. \ref{betadef} over a wide range of $L/\lambda$ and $\omega\tau$ values and with strong agreement in the quadratic scaling regime ($L/\lambda>\pi$).}
    \label{betaplot}
\end{figure}

We can obtain a numerically derived estimate of the response scaling function by performing a linear fit to the data presented in the insets of Fig. \ref{SVresp}. Fig. \ref{betaplot} compares these fitted values to Eq. \ref{betadef} for a wide range of $L/\lambda$ and $\omega\tau$ values with good agreement. Since any boundary condition shift can be decomposed into a Fourier sum over a set of frequencies and we have shown $\beta(L/\lambda)$ to be the response scaling function for all frequencies, we can thus extrapolate $\tau\beta(L/\lambda)$ to be the response time scale given any arbitrary shift in boundary conditions for a single vessel.

\subsection{Network Mechanics}
\label{sec:networks}

We now consider a multitude of compliant, fluid carrying vessels interconnected to form a fluid transport network of nodes and edges. For bookkeeping purposes and without loss of generality we can assign a directionality to each edge, e.g. edge $e=( \mu, \nu)$ is traversed from $\mu$ to $\nu$. We identify the location along each vessel with the variable $z$, which is $z=0$ at the node $\mu$ where the edge is outgoing from, and $z=L_{\mu\nu}$ where the edge is incoming to. Each individual vessel is still considered to obey Eqs. \ref{Qsym} and \ref{Psym}, but each vessel may have its own unique values for $\lambda$, $\tau$, and $\alpha$, which are $z$ independent. Specifically, we identify $\lambda_{\mu\nu}$ as the value of $\lambda$ within the vessel that begins at network node $\mu$ and ends at node $\nu$ with the same index notation also being applied to all other parameters and variables. Scalar quantities such as the characteristic length scale $\lambda_{\mu\nu}$ are independent of the direction traversed between the nodes and thus symmetric with respect to an interchange of indices. For spatially dependent quantities such as the pressure $P_{\mu\nu}(z,t)$, the order of the indices implies the directionality of the edge and $P_{\mu\nu}(z,t) = P_{\nu\mu}(L_{\mu\nu}-z,t)$, so that $P_{\mu\nu}(0,t)$ can always be identified with the pressure $P_{\mu}(t)$ at node $\mu$. Quantities which depend on the direction the edge is traversed such as $Q_{\mu\nu}(z,t)$ change sign when the beginning and ending nodes are switched and are thus antisymmetric with respect to an interchange of indices, so that $Q_{\mu\nu}(z,t) =  - Q_{\nu\mu}(L_{\mu\nu}-z,t) $. 

\begin{figure}[t]
    \centering
    \includegraphics[width=\columnwidth]{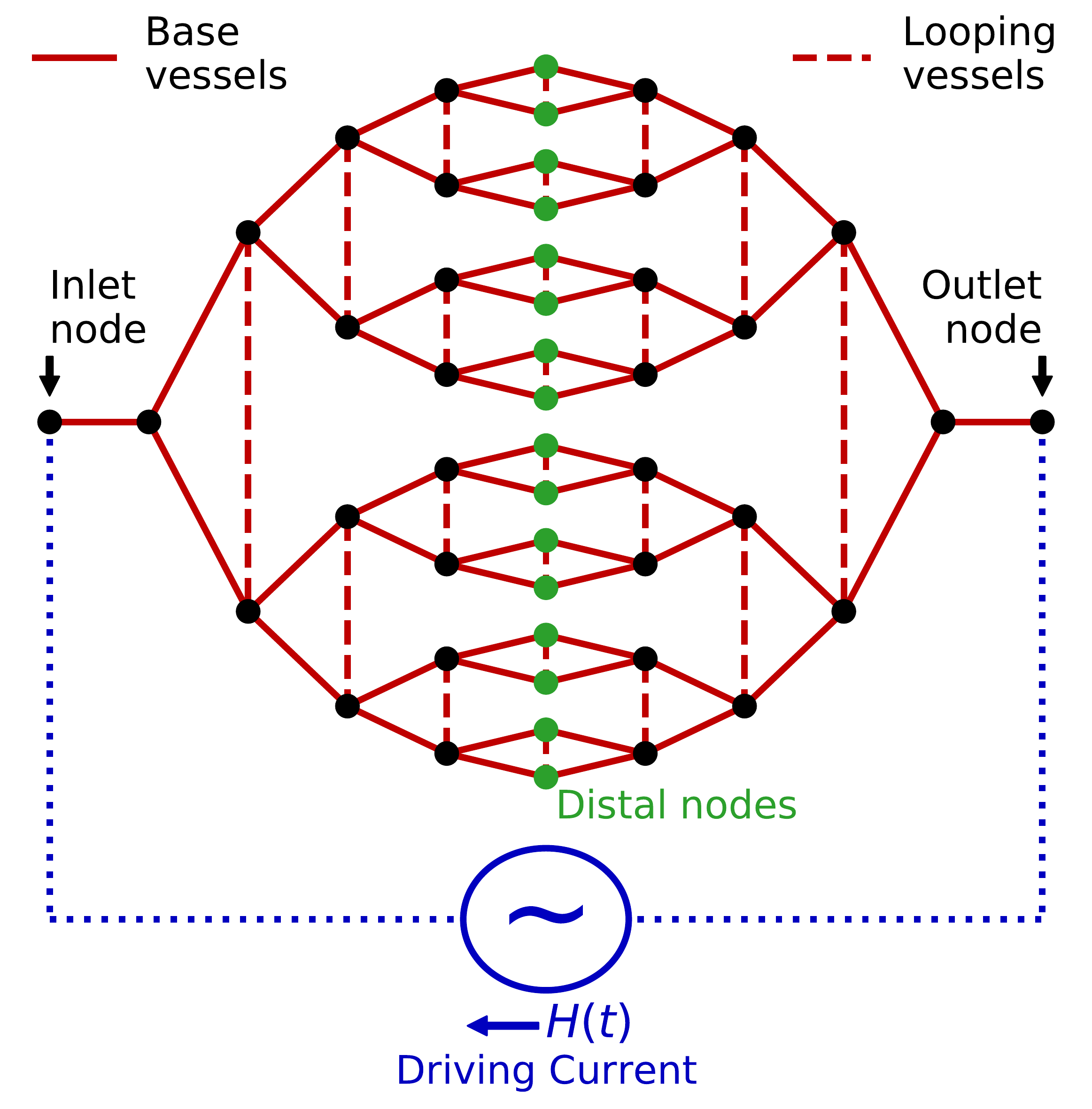}
    \caption{Diagram of toy network structure. Vessels represented by red lines connect at nodes represented by dots. Solid lines show base vessels that connect the inlet and outlet nodes to the distal nodes (green dots), and dashed lines show looping vessels that allow fluid to flow between branching generations. A current driver (blue) provides an arbitrary externally imposed current $H(t)$ into the inlet node and out of the outlet node via external vessels (dotted line).}
    \label{netdiag}
\end{figure}

The connectivity laws of the network are taken to be two-fold: 1) pressure is continuous across networks nodes and 2) the total current being inputted into a node must equal the total current flowing away from it through the network. These are manifested mathematically by enforcing that

\begin{equation}
    P_{\mu\nu}\left(0,t\right) = P_{\mu}\left(t\right) \quad \forall \quad \nu\in\mathcal{N}_{\mu},
    \label{Pconnect}
\end{equation}

\begin{equation}
    \sum_{\nu\in\mathcal{N}_{\mu}}Q_{\mu\nu}(0,t) = H_{\mu}\left(t\right),
    \label{Qconnect}
\end{equation}

\noindent where $H_{\mu}(t)$ is the current being inputted into node $\mu$ by an external source and $\mathcal{N}_{\mu}$ is the set of all nodes connected to node $\mu$ by a single vessel. For example, the network depicted in Fig. \ref{netdiag} has $H_{\mu}(t)=0$ for all internal nodes while $H_{\text{inlet}}=H(t)$ and $H_{\text{outlet}}=-H(t)$. Thus, for each node there are two possible boundary conditions to specify, $P_{\mu}(t)$ and $H_{\mu}(t)$, creating a total of $2N$ possible boundary conditions to specify for the whole network, where $N$ is the number of nodes.

However, there exists and interdependence between the set of $P_{\mu}(t)$ and $H_{\mu}(t)$ that reduces the necessary number of specified boundary conditions to a subset of size $N$. This can be shown by considering the relationship between $P_{\mu}(t)$ and $H_{\mu}(t)$. By expanding $P_{\mu\nu}(z,t)$ and $Q_{\mu\nu}(z,t)$ into their respective Fourier series in time, Eqs. \ref{Qsym} and \ref{Psym} can be solved to express $\tilde{P}_{\mu\nu}^{(n)}(z)$ and $\tilde{Q}_{\mu\nu}^{(n)}(z)$, the Fourier transformed vessel pressures and currents, as linear combinations of $\tilde{P}_{\mu}^{(n)}$ and $\tilde{P}_{\nu}^{(n)}$, the Fourier transformed node pressures (Eqs. \ref{FTQsol_Pbound} and \ref{FTPsol_Pbound}). This in turn allows the second connectivity law to be rewritten as $\tilde{H}_{\mu}^{(n)}=\sum_{\nu}\mathcal{L}_{\mu\nu}^{(n)}\tilde{P}_{\nu}^{(n)}$, where $\mathcal{L}^{(n)}$ is the network Laplacian matrix in Fourier space and is given by Eq. \ref{Lmndef}. This creates a set of $N$ linearly independent equations for each nonzero frequency. For the zero frequency mode the matrix $\mathcal{L}^{(0)}$ only has rank $N-1$ as a choice of gauge must be made to fully determine all $\tilde{P}_{\mu}^{(0)}$. Thus, once a gauge is defined the space of undetermined variables per frequency is reduced from the full $2N$ values of $\tilde{P}_{\mu}^{(n)}$ and $\tilde{H}_{\mu}^{(n)}$ down to a subset of only $N$ values.

With the mechanics and necessary boundary conditions defined, we now construct a simple toy network to test how its properties compare to those derived for the single vessel. Fig. \ref{netdiag} depicts a hierarchical network designed to be reminiscent of a small, idealized vascular network in which the input and output vessels branch inward over several generations to meet at a central set of distal nodes. Solid lines represent the base vessels that allow for fluid to reach each of the distal nodes while dashed lines represent looping vessels that allow for greater mixing of the fluid flow. For now we consider a relatively simple case with no looping vessels in which all remaining vessels are of equal length and each parent vessel branches into two daughter vessels each with their own values of $\lambda$, $\tau$, and $\alpha$. The size of the daughter vessels are taken to obey

\begin{equation}
    a_{\text{parent}}^{\gamma} = ba_{\text{daughter }1}^{\gamma}+\left(1-b\right)a_{\text{daughter }2}^{\gamma},
    \label{branchingeq}
\end{equation}

\noindent where each $a$ in Eq. \ref{branchingeq} is the radius of the cross section of the respective vessel, $\gamma$ is the branching exponent, and $b$ is the branching ratio. Thus, $\gamma=2$ corresponds to branching in which the total cross sectional area is preserved. How the cross sectional area of the daughters compares to that of the parent is relevant due to $\lambda$, $\tau$, and $\alpha$ each scaling linearly with this area when the properties of the fluid and vessel wall material are held fixed (see Appendix \ref{LNSE}).

Of important note is that we make no explicit enforcement of impedance matching at the bifurcation points. In biological contexts, impedance matching has the effect of reducing energy dissipation by minimizing wave reflection and is thus typically imposed on the basis of minimal dissipation being advantageous. In Fig. \ref{netresp}A we will consider the biologically relevant cases of networks that obey Eq. \ref{branchingeq} with $\gamma=2$ or $3$, each of which satisfy impedance matching in different regimes of the vascular network \cite{savage2008sizing,hughes2015optimality}. However, as we are primarily interested in response time as opposed to dissipation we also consider a class of networks which do not obey Eq. \ref{branchingeq} and thus do not satisfy impedance matching, as will be represented in Fig. \ref{netresp}B.

\begin{figure}[htp]
    \centering
    \includegraphics[width=\columnwidth]{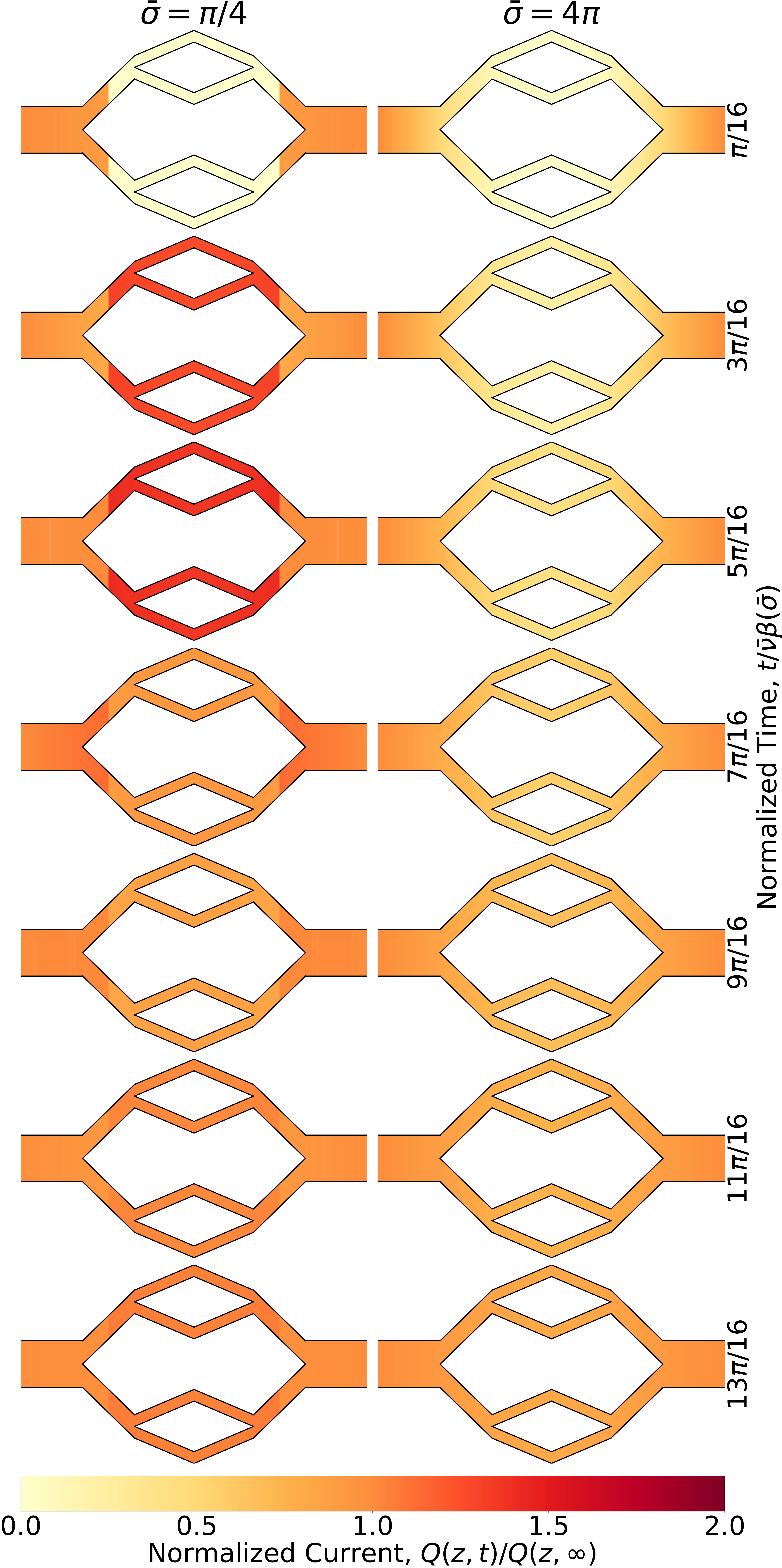}
    \caption{Example of current distribution in a smaller, two generation version of the toy network depicted in Fig. \ref{netdiag} for two distinct values of $\bar{\sigma}$. In each case, the branching exponent and branching ratio are taken to be $\gamma=2$ and $b=1/2$. Current at each location within the networks is normalized by the long time steady state value, $Q(z,\infty)$ while time is normalized by $\bar{\nu}$ and $\beta(\bar{\sigma})$ as defined in Eq. \ref{betadef}. Similar to the single vessel case presented in Fig. \ref{Qframes}, the smaller ($\bar{\sigma}=\pi/4$) network shows reflecting wavefronts that cause the current to undergo decaying oscillations while the larger ($\bar{\sigma}=4\pi$) network shows a gradual approach to steady state over a longer timescale. The depicted current dynamics are also shown in supplemental videos.}
    \label{netframes}
\end{figure}

We now investigate the properties of such networks compared to those of the single vessel derived previously. To begin, we note that in a single vessel any wavefront that isn't part of a perfectly periodic signal will travel a distance $z$ in a time $z\tau/\lambda$ and decay as $\text{exp}(-z/\lambda)$. In a network, such wavefronts will necessarily split when they encounter branching nodes. Since these wavefronts travel with velocity $\lambda/\tau$ within each individual vessel, the time to travel from one position to another along a specific path $\mathcal{S}$ through the network can be expressed as $t_{\mathcal{S}}=\int_{\mathcal{S}}dz\>\tau(z)/\lambda(z)$, where the integral is over the path $\mathcal{S}$. Similarly, the wavefront will decay as $\text{exp}(-\int_{\mathcal{S}}dz\> 1/\lambda(z))$. For a network such as the one shown in Fig. \ref{netdiag}, we can use this decay function to generalize the expression $L/\lambda$, the nondimensionalized size of the single vessel, to the path dependent $\sigma_{\mathcal{S}} = \int_{\mathcal{S}} dz\> 1/\lambda(z)$, where $\mathcal{S}$ is any path from the far left node to the far right node that does not backtrack. This gives us a spatial scale of a single path to use in the same way $L/\lambda$ was used for the results presented in Fig. \ref{SVresp}. To obtain a similar scale for the entire network, we perform a weighted average of $\sigma_{\mathcal{S}}$ over all possible paths in which the weight of each path is the current that runs through that particular path when a steady, nonpulsatile flow in inputted into the far left node and outputted out of the far right node. We denote this current averaged value as $\bar{\sigma}$. We can further obtain a temporal scale, $\nu_{\mathcal{S}}$, for a single path by considering the time required for a wavefront to traverse the path normalized by the $\sigma_{\mathcal{S}}$ value of that path, $\nu_{\mathcal{S}}=(\int_{\mathcal{S}} dz\>\tau(z)/\lambda(z))/\sigma_{\mathcal{S}}$. This is equivalent to defining $\tau$ by the relation $\tau=(z\tau/\lambda)/(z/\lambda)$ in the single vessel case. The current averaged $\bar{\nu}$ can then be calculated using the same weighting scheme as $\bar{\sigma}$.

Fig. \ref{netframes} shows how $\bar{\sigma}$ and $\bar{\nu}$ are analogous to the quantities $L/\lambda$ and $\tau$ from the single vessel case. By considering a smaller version of the toy network shown in Fig. \ref{netdiag} with only two branching generations, no looping vessels, and fixed values of the branching exponent ($\gamma=2$) and branching ratio ($b=1/2$), we see that the two systems represented in Figs. \ref{Qframes} and \ref{netframes} are qualitatively equivalent. In both cases, the smaller system ($L/\lambda$ and $\bar{\sigma}=\pi/4$) exhibits reflecting wavefronts that cause the current to undergo decaying oscillations while the larger system ($L/\lambda$ and $\bar{\sigma}=4\pi$) simply approaches its long time steady state gradually.

We can now numerically calculate the response scaling function of the network in the same way as was done for the single vessel in Fig. \ref{betaplot}. We consider step function boundary conditions for the current in the inlet and outlet nodes, similar to those used to produce the $\omega=0$ case of Eq. \ref{Qsymsol}. Specifically, a current of the form $\Theta(t)$ is inputted into the inlet node and outputted out of the outlet node. We neglect the case of pulsatile boundary conditions as it was shown in the single vessel case to produce identical response times. The system is simulated by numerically evolving it through discretized versions of Eqs. \ref{Qsym} and \ref{Psym}. We monitor the total current passing through all the central nodes and denote the residual current as this total central current subtracted from the long time limit. We then perform a linear fit to the logarithm of the magnitude of the residual current, as was done for the data presented in Fig. \ref{SVresp}, to obtain a measure of the response scaling function as a function of the network size, $\beta(\bar{\sigma})$. To perform this fit, time is rescaled by $\bar{\nu}$, thus implying that the total response time of the network is $\bar{\nu}\beta(\bar{\sigma})$.

\begin{figure}[t]
    \centering
    \includegraphics[width=\columnwidth]{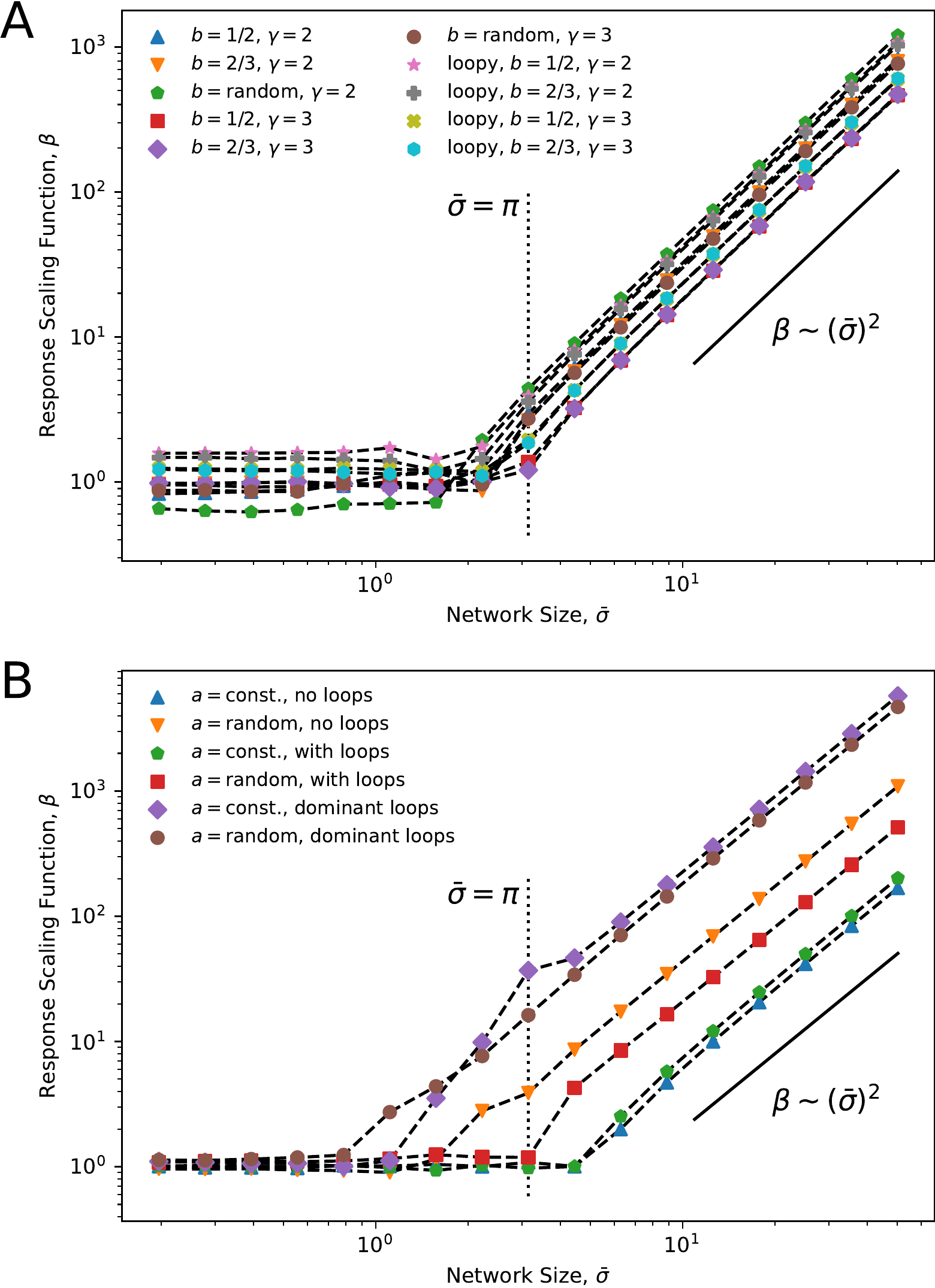}
    \caption{Response scaling function, $\beta$, as a function of network size, $\bar{\sigma}$, for a variety of different networks with topology presented in Fig. \ref{netdiag}. A) Hierarchical networks that obey Eq. \ref{branchingeq} show constant and quadratic scaling regimes, as in the single vessel case, with critical values of $\bar{\sigma}$ consistently less than $\pi$. The tight grouping of the 10 plots shown indicates that $\beta$ as a function of $\bar{\sigma}$ is very weakly dependent on the choice $b$ and $\gamma$ when time is measured in terms of $\bar{\nu}$. B) Homogenous networks with all vessel sizes drawn from the same distribution also show constant and quadratic scaling regimes with a far broader range of critical $\bar{\sigma}$ values.}
    \label{netresp}
\end{figure}

Fig. \ref{netresp} shows the numerically measured value of $\beta$ as a function of network size, $\bar{\sigma}$ for two different classes of networks. In Fig. \ref{netresp}A, Eq. \ref{branchingeq} is used to construct the network depicted in Fig. \ref{netdiag} for a variety of different $b$ and $\gamma$ values, both with and without loops and including cases in which $b$ is chosen randomly from a symmetric triangular distribution between $0.05$ and $0.95$ with independent draws at each branching node. As can be seen from Fig. \ref{netresp}A, the response scaling function for these networks has many of the same qualitative features as that for the single vessel seen in Fig. \ref{SVresp}. $\beta$ holds a constant value near $1$ for small $\bar{\sigma}$ and scales quadratically with $\bar{\sigma}$ for large $\bar{\sigma}$. One important distinction is the location of the critical value. In the single vessel, $L/\lambda=\pi$ was the critical value that marked the transition from the constant to quadratic regimes, but Fig. \ref{netresp}A shows that this transition occurs at $\bar{\sigma}<\pi$ and is not constant across all networks. Additionally, the networks tend to separate into different groups depending on which side of this critical value they are on. For small $\bar{\sigma}$ we notice that the networks separate into those with loops and those without, the former maintaining $\beta>1$ and the later maintaining $\beta<1$. Conversely, for large $\bar{\sigma}$ the ordering criteria change so that the $\gamma=3$ networks maintain a smaller value of $\beta$ than the $\gamma=2$ networks. The difference in values of $\beta$ between these groups is small, however, indicating that $\beta$ depends on $b$ and $\gamma$ very weakly.

We also considered a class of networks with the same topology, but different determination of the vessel sizes. To determine the vessel radius, $a$, we considered one set of cases in which all vessels have identical and constant radii and one set in which the radius of each vessel is determined randomly and independently as $a=2^{y}$, where $y$ is a standard normal random variable. Additionally, we simulated one pair of networks (one network with constant $a$ and one with random $a$) with no looping vessels, one pair with looping vessels with radius determined in a manner identical to the nonlooping vessels, and one pair in which the looping vessels were made dominant by increasing their radii by a factor of $10$ after they were again determined in a manner identical to the nonlooping vessels. The numerically measured values of $\beta$ for each of these networks at a variety of different $\bar{\sigma}$ values are shown in Fig. \ref{netresp}B. Once again we see that for sufficiently small $\bar{\sigma}$ the response scaling function maintains a constant value of very nearly $1$ for all such networks, while for large $\bar{\sigma}$ it scales quadratically with $\bar{\sigma}$. The critical values of $\bar{\sigma}$ where $\beta$ transitions from being constant to quadratic are seen to exist over a much larger range that extend above $\pi$ for some networks, in contrast to those seen in Fig. \ref{netresp}A where all the critical $\bar{\sigma}$ values were smaller than $\pi$.

\section{Discussion}
\label{sec:disc}

We have shown that fluid flow through a single cylindrical vessel comprised of compliant walls obeys Eqs. \ref{Qeq} and \ref{Peq} in the linear regime. By analyzing these equations in terms of $W(z,t)$ and Eq. \ref{Wsym}, we were able to derive the response scaling function, $\beta(L/\lambda)$, for such a vessel under symmetric boundary conditions. This function showed that short vessels ($L/\lambda<\pi$) are dominated by reflecting wavefronts that decay at a rate independent of the vessel size while long vessels ($L/\lambda>\pi$) behave as a diffusive medium and respond over a time that scales quadratically with the vessel size. Generalizing our theory to a network of such vessels and numerically calculating the response scaling function, we see that these two specific regimes exist for a wide variety of different branching networks, though the critical $\bar{\sigma}$ value that separates them varies between the networks presented in Fig. \ref{netresp}. This implies that a branching network such as the one depicted in Fig. \ref{netdiag} has a response scaling function that is similar in form to $\beta(\bar{\sigma})$ as defined in Eq. \ref{betadef} but with the critical $\bar{\sigma}$ value that separates the constant $\beta=1$ regime from the quadratic $\beta\sim\bar{\sigma}^{2}$ regime shifted by an amount that varies between the networks considered here.

The response scaling function has also been shown to dictate the response time not just for the mean of the current but the pulsatile components as well. This has an important consequence in that for any given network of compliant vessels there is a set timescale over which any section of the network will be able to respond to changes in any other section. Specifically, the time for any general wavefront to traverse a path $\mathcal{S}$ is simply given by $\sigma_{\mathcal{S}}\nu_{\mathcal{S}}=\int_{\mathcal{S}} dz\>\tau(z)/\lambda(z)$, but if the path is too long relative to the values of $\lambda$ found along that path then the resulting wavefront will be substantially decayed and the time required to see a significant change in the current or pressure will be increased above the value $\sigma_{\mathcal{S}}\nu_{\mathcal{S}}$. Thus, the timescale $\sigma_{\mathcal{S}}\nu_{\mathcal{S}}$ sets a limit on how quickly mechanical information in the form of fluid wavefronts can be transmitted along that path. In order for fluid pressure and flow to respond more quickly, external information transmission is necessary. In biological contexts, this can be achieved through electrical signals in the nervous system, and the values of $\sigma_{\mathcal{S}}$ and $\nu_{\mathcal{S}}$ along certain paths may dictate where in the body such electrical information transmission is most necessary to maintain proper blood flow in response to sudden changes such as shifts in gravity from different body positions. The transmission of mechanical information in biological networks has some conceptual similarity to the propagation of the effects of a link failure in power grids \cite{Schafer2018,Zhang2020} - a concrete investigation of the analogies could help understand more subtle aspects of the effects of topology on the response of vascular networks to mechanical perturbations. 

The theory presented here captures the qualitative behaviors of a network of compliant vessels transmitting pulses in flow and pressure, but it has limitations related to the various approximations made to derive Eqs. \ref{Qsym} and \ref{Psym}. Therefore, we do not expect a strict quantitative agreement in matters regarding the exact propagated waveform shape or the transmission and reflection coefficients at the nodes. Two distinct instances of such limitations come from assumptions made during the handling of $-\nabla^{2}u_{z}$. The axial term was neglected completely as we assumed the wavelength of the waves traversing the system would be significantly larger than the vessel radius. While calculations of wavelength from pulse wave velocity and driving frequency in the human vasculature typically show this to be valid \cite{pan2014one,alastruey2011pulse}, any system in which this is not valid would require a fourth parameter beyond $r$, $\ell$, and $c$ or $\lambda$, $\tau$, and $\alpha$ as well as a more complicated form of the function $k(\omega\tau)$. Additionally, the radial term was simplified under the assumption that laminar flow with a quadratic velocity profile is perpetually established within the vessel. This is not true in general, especially in a biological context, where the Womersley number can range from order $\sim 10$ in the major blood vessels, implying the flow oscillates too rapidly to maintain a quadratic velocity profile, to order $\sim 10^{-2}$ in the minor blood vessels, implying a quadratic velocity profile can be maintained \cite{womersley1955method,west1997general}. For nonquadratic profiles, the resistive term of Eq. \ref{Peq} would have to be reevaluated based on the alternate profile used. However, for the human vasculature specifically, the quadratic approximation is only violated in the largest vessels and is thus valid for the majority of the network.

Possibly the most significant limitation of this theory are the linear approximations. In the single vessel case, the nonlinear term of Eq. \ref{NSeq} is neglected. Previous computational studies have shown that nonlinear models of fluid flow through compliant vessels are better able to reproduce experimentally measured pressure waveforms in the major arteries \cite{bui2009dynamics,alastruey2011pulse}. The model presented here can be made to reflect these nonlinearities by reincorporating the neglected terms from Eq. \ref{NSeq}. In the network case, another possible source of nonlinearity is in the pressure connectivity law, Eq. \ref{Pconnect}. This can be made to more accurately incorporate Bernoulli's principle by enforcing that the sum of the pressure and kinetic energy density is continuous across a network node rather than just the pressure. It will be interesting to explore how these nonlinearities affect response time in future works.

Despite these limitations, which are present in the majority of vascular models that also linearize the flow equations \cite{sherwin2003one,alastruey2012physical,cousins2013new,flores2016novel,yigit2016non}, the theory presented here provides many valuable insights. In particular, the values of $\bar{\sigma}$ and $\bar{\nu}$ can help determine the behavior of complex networks for which data is available. As an example, we consider the human vascular system. We can estimate the aorta to have a distensibility of $8.9\times 10^{-3}\>\text{mm Hg}^{-1}$ and cross sectional area of $515\>\text{mm}^{2}$ \cite{voges2012normal}, while the blood in the aorta has a dynamic viscosity of $3.5\times 10^{-3}\>\text{Pa}\cdot\text{s}$ and density of $1050\>\text{kg}\cdot\text{m}^{-3}$ \cite{kumar2017non}. Using these values we can calculate the resistance and inertia per unit length via the aforementioned relations $r=8\pi\mu/A_{0}^{2}$ and $\ell=\rho/A_{0}$ and the compliance per unit length as $c=DA_{0}$, where $D$ is the distensibility. This particular formula for the compliance can be easily obtained from the definition of distensibility presented in \cite{voges2012normal} and the assumption that changes in cross sectional area are small and obey $\Delta A=c\Delta P$. These values of $r$, $\ell$, and $c$ then allows us to derive the values $\lambda\approx 54.5\>\text{m}$ and $\tau\approx 14.5\>\text{s}$ from the aforementioned relations $\lambda=2\sqrt{\ell/c}/r$ and $\tau=2\ell/r$. We can use these values of $\lambda$ and $\tau$ in conjunction with $A_{0}\approx 515\>\text{mm}^{2}$ and an assumed heartrate of 75 bpm to obtain $\lambda/a_{0}\approx 4260$ and $\abs{k(\omega\tau)}\approx 114$, thus verifying the condition $\abs{k(\omega\tau)}\ll\lambda/a_{0}$. We can also verify that the assumption that the pulse velocity, $\lambda/\tau\approx 3.76\>\text{m/s}$, is notably larger than the maximum blood flow velocity of approximately $120\>\text{cm/s}$ \cite{voges2012normal}. Under the assumption that $\lambda$ and $\tau$ each scale linearly with $A_{0}$, these approximations should hold throughout the rest of the vasculature as well (see Appendix \ref{LNSE} for discussion of these conditions and scalings).

From here we consider a hypothetical vessel that begins with a cross sectional area equivalent to that of the aorta and over a length of $1\>\text{m}$ tapers down in such as way that the area linearly decreases to a value of $50\>\mu\text{m}^{2}$. The vessel is then considered to loop back to its starting point in a symmetric way so that its area linearly increases from $50\>\mu\text{m}^{2}$ back to $515\>\text{mm}^{2}$ over a second length of $1\>\text{m}$. By making the simple assumptions that $\lambda$ and $\tau$ will scale linearly with area (see Appendix \ref{LNSE}), we can calculate that $\sigma_{\mathcal{S}}=\int_{\mathcal{S}}dz\>1/\lambda(z)\approx 0.6$ and $\nu_{\mathcal{S}}=(\int_{\mathcal{S}}dz\>\tau(z)/\lambda(z))/\sigma_{\mathcal{S}}\approx 0.9\>\text{s}$ for the path that loops from one end of this vessel to the other. By assuming that this distance of $1\>\text{m}$ is a good approximation of the average distance blood must travel to get from the heart to a capillary bed, we can take this hypothetical vessel to be a very rough estimate of a characteristic path in the human vasculature in the sense that the network averaged values of $\sigma$ and $\nu$ are equal to those of this particular path ($\bar{\sigma}\approx\sigma_{\mathcal{S}}$ and $\bar{\nu}\approx\nu_{\mathcal{S}}$).

Specifically, by comparing this value of $\bar{\sigma}$ to those seen within Fig. \ref{netresp}, we can extrapolate that the human vascular network is likely within the constant response regime rather than the quadratic regime, but near the transition point. Additionally, the value of $\bar{\nu}$ dictates that the response time in this region is on the order of $1\>\text{s}$, in agreement with timescales found in rats for changes in local oxygen concentration after a shift in heart rate occurs \cite{masamoto2007apparent}. While rats are much smaller than the $1\>\text{m}$ distance used to obtain the estimated value of $\bar{\nu}$, $\nu_{\mathcal{S}}$ along any single path is only weakly dependent on distance due to it being normalized by $\sigma_{\mathcal{S}}$, thus allowing an approximate comparison to be made. From these findings, we can make the prediction that the vascular network may be adapted to restrict itself to the region of minimal possible response time. More detailed measurements of the value of $\bar{\sigma}$ and $\bar{\nu}$ within the vasculature of humans as well as other animals are needed to verify this possibility and may reveal that existing at or near this transition point is a universal trend. Moreover, they may shed light on situations where the body appears to have evolved to harness pulsatility to perform specific functions, such as the movement of cerebrospinal fluid \cite{Mestre2018}.

By examining the single vessel case, we can understand several observable effects of existing at this transition point. In single vessels at the critical value of $L/\lambda=\pi$, the wavefronts traveling through the vessel decay enough that the reflections do not cause sudden spikes in flow and dissipation, such as those seen in the blue and green curves of Fig. \ref{SVresp}, but are not so decayed as to enter the regime in which flow and pressure expand diffusively. We can extrapolate these findings to networks to predict that when $\bar{\sigma}$ is too small wave reflections become highly significant and large spikes in flow and pressure should be visible. This phenomenon is seen in the effects of arterial stiffening. As blood vessels become stiffer, or equivalently their compliance lowers, the value of $\lambda$ within each vessel must increase, in turn causing an increase in pulse wave velocity ($\lambda/\tau$) and decrease in $\bar{\sigma}$. This increase in wave velocity as well as increased amplitude of reflected waves can be directly observed in hypertensive patients with increased arterial stiffness \cite{weber2004arterial,nichols2008effects}.

Conversely, our theory predicts that should arterial stiffness be lowered not only would wave velocity and reflected wave amplitude also decrease, but if the increased compliance causes the $\bar{\sigma}$ value of the network to move past the transition point then a marked increase would also occur in the time required to establish a change in blood flow in the capillary bed. Moving past this transition point could be particularly detrimental for networks with significant loops due to the significantly smaller critical $\bar{\sigma}$ value seen in the purple and brown curves of Fig. \ref{netresp}B. However, since arterial stiffness tends to increase rather than decrease as a consequence of age and/or disease, this prediction is far more difficult to verify with existing measurements. A more complete understanding of how these effects might apply to the human vascular network could represent a step towards being able to better diagnose disease and construct prosthetics and cardiac aids that work natively with existing blood vessels.

\acknowledgements This research was supported by the NSF Award PHY-1554887 and the Simons Foundation through Award 568888.

\bibliography{refs}
\bibliographystyle{ieeetr}

\appendix
\section{Linearizing the Navier-Stokes Equation}
\label{LNSE}

Following the techniques described in \cite{van1998cardiovascular}, here we derive Eqs. \ref{Qeq}-\ref{Psym} from Eqs. \ref{incom} and \ref{NSeq}. We begin by making two important assumptions. The first is rotational symmetry, meaning that all dynamic variables must be independent of angular position within the cylindrical vessel and the angular flow velocity must be 0. The second is that the fluid is incompressible, meaning the flow velocity must obey Eq \ref{incom}. From here we define the volumetric flow rate

\begin{equation}
Q\left(z,t\right) = \int dA\> u_{z}\left(z,r,t\right) = 2\pi\int_{0}^{a}dr\>ru_{z}\left(z,r,t\right),
\label{Qurel}
\end{equation}

\noindent where $\int dA$ represents integration over the circular cross section and $a$ is the radius of the cross section at axial position $z$. Integrating Eq. \ref{incom} over the cross sectional area thus yields

\begin{align}
0 &= \int dA\left(\frac{\partial u_{z}}{\partial z}+\frac{1}{r}\frac{\partial}{\partial r}\left(ru_{r}\right)\right) \nonumber\\
&= \frac{\partial Q}{\partial z}+2\pi\int_{0}^{a}dr\>\frac{\partial}{\partial r}\left(ru_{r}\right) = \frac{\partial Q}{\partial z}+2\pi au_{r}\left(z,a,t\right).
\label{udivint}
\end{align}

\noindent Finally, we note that if the radial velocity at $r=a$ is nonzero, then $a$ itself must be changing at the same rate in order to accommodate the expanding or contracting fluid. This can be expressed as

\begin{equation}
\frac{\partial A}{\partial t} = \frac{\partial}{\partial t}\left(\pi a^{2}\right) = 2\pi a\frac{\partial a}{\partial t} = 2\pi au_{r}\left(z,a,t\right).
\label{dAdt}
\end{equation}

\noindent Inserting Eq. \ref{dAdt} into Eq. \ref{udivint} produces

\begin{equation}
\frac{\partial Q}{\partial z}+\frac{\partial A}{\partial t} = \frac{\partial Q}{\partial z}+\frac{\partial A}{\partial P}\frac{\partial P}{\partial t} = 0.
\label{QArel}
\end{equation}

\noindent Thus, we can see that Eq. \ref{Qeq} can be produced solely via the assumptions that the fluid is rotationally symmetric and incompressible.

Next we turn to the Navier-Stokes equation itself, which for an incompressible fluid and no external forces, can be written as Eq. \ref{NSeq}. By isolating the nonlinear term, $(\vec{u}\cdot\vec{\nabla})\vec{u}$, on the right hand side and expanding the $\vec{\nabla}$ operator into its axial and radial parts (angular parts are ignored due to the assumption of rotational symmetry), the axial component of Eq. \ref{NSeq} can be extracted in the form

\begin{align}
&\frac{\partial p}{\partial z}+\rho\frac{\partial u_{z}}{\partial t}-\mu\left(\frac{\partial^{2}u_{z}}{\partial z^{2}}+\frac{1}{r}\frac{\partial}{\partial r}\left(r\frac{\partial u_{z}}{\partial r}\right)\right) \nonumber\\
&= -\rho\left(u_{r}\frac{\partial u_{z}}{\partial r}+u_{z}\frac{\partial u_{z}}{\partial z}\right).
\label{NavierStokesaxial}
\end{align}

\noindent We then average Eq. \ref{NavierStokesaxial} over the cross section area to produce

\begin{align}
&\frac{1}{A}\int dA\left(\frac{\partial p}{\partial z}+\rho\frac{\partial u_{z}}{\partial t}-\mu\frac{\partial^{2}u_{z}}{\partial z^{2}}-\mu\frac{1}{r}\frac{\partial}{\partial r}\left(r\frac{\partial u_{z}}{\partial r}\right)\right) \nonumber\\
&= \frac{\partial P}{\partial z}+\frac{\rho}{A}\frac{\partial Q}{\partial t}-\frac{\mu}{A}\frac{\partial^{2}Q}{\partial z^{2}}-\frac{2\pi\mu}{A}a\left.\frac{\partial u_{z}}{\partial r}\right|_{r=a} \nonumber\\
&= -\frac{\rho}{A}\int dA\left(u_{r}\frac{\partial u_{z}}{\partial r}+u_{z}\frac{\partial u_{z}}{\partial z}\right),
\label{NSaxialave}
\end{align}

\noindent where the area averaged pressure, $P$, is defined as

\begin{equation}
P\left(z,t\right) = \frac{1}{A}\int dA\>p\left(z,r,t\right).
\label{Pprel}
\end{equation}

From here we assume that the fluid is Newtonian and changes in the flow happen over relatively long time scales, thus allowing the flow to be approximately fully developed at all times. This means $u_{z}$ must follow the Hagen-Poiseuille equation and be expressable as

\begin{equation}
u_{z}\left(z,r,t\right) = U\left(z,t\right)\left(1-\frac{r^{2}}{a^{2}}\right).
\label{HPuz}
\end{equation}

\noindent Assuming Eq. \ref{HPuz} is equivalent to restricting the Womersely number of the system to be small in the case of pulsatile flow. Inserting Eq. \ref{HPuz} into Eq. \ref{Qurel} then yields

\begin{align}
Q\left(z,t\right) &= 2\pi\int_{0}^{a}dr\>rU\left(z,t\right)\left(1-\frac{r^{2}}{a^{2}}\right) \nonumber\\
&= \frac{1}{2}\pi a^{2}U\left(z,t\right) = \frac{1}{2}AU\left(z,t\right).
\label{HPQu}
\end{align}

\noindent We can also differentiate Eq. \ref{HPuz} and combine it with Eq. \ref{HPQu} to produce

\begin{equation}
-a\left.\frac{\partial u_{z}}{\partial r}\right|_{r=a} = 2U\left(z,t\right) = \frac{4}{A}Q.
\label{HPuzdif}
\end{equation}

Turning to the nonlinear terms on the right hand side of Eq. \ref{NSaxialave}, we first note that Eq. \ref{HPuz} forces $u_{z}$ to vanish at $r=a$. Additionally, rotational symmetry requires that $u_{r}$ vanish at $r=0$. Using these vanishing boundary conditions along with Eq. \ref{incom} allows us to remove $u_{r}$ from Eq. \ref{NSaxialave} via the relation

\begin{align}
&\frac{\rho}{A}\int dA\left(u_{r}\frac{\partial u_{z}}{\partial r}\right) = -\frac{2\pi\rho}{A}\int_{0}^{a}dr\>u_{z}\frac{\partial}{\partial r}\left(ru_{r}\right) \nonumber\\
&= \frac{2\pi\rho}{A}\int_{0}^{a}dr\>u_{z}r\frac{\partial u_{z}}{\partial z} = \frac{\rho}{A}\int dA\>u_{z}\frac{\partial u_{z}}{\partial z}.
\label{urint}
\end{align}

\noindent Eq. \ref{urint} shows that the two nonlinear terms in Eq. \ref{NSaxialave} are equivalent. With is, we can use Eqs. \ref{HPuz} and \ref{HPQu} to express the nonlinear portion of Eq. \ref{NSaxialave} as

\begin{align}
&\frac{\rho}{A}\int dA\>2u_{z}\frac{\partial u_{z}}{\partial z} = \frac{\rho}{A}\frac{\partial}{\partial z}\int dA\left(U\left(1-\frac{r^{2}}{a^{2}}\right)\right)^{2} \nonumber\\
&= \frac{\rho}{A}\frac{\partial}{\partial z}\left(U^{2}\frac{\pi a^{2}}{3}\right) = \frac{4\rho}{3A}\frac{\partial}{\partial z}\left(\frac{Q^{2}}{A}\right) \nonumber\\
&= \frac{4\rho Q}{3A}\left(\frac{2}{A}\frac{\partial Q}{\partial z}-\frac{Q}{A^{2}}\frac{\partial A}{\partial z}\right).
\label{nonlinint}
\end{align}

\noindent Inserting Eq. \ref{HPuzdif} and \ref{nonlinint} into Eq. \ref{NSaxialave} and expressing $A$ as a function of $P$, as was done in Eq. \ref{QArel}, yields

\begin{align}
&\frac{\partial P}{\partial z}+\frac{\rho}{A}\frac{\partial Q}{\partial t}+\frac{8\pi\mu}{A^{2}}\left(Q-\frac{A}{8\pi}\frac{\partial^{2}Q}{\partial z^{2}}\right) \nonumber\\
&= -\frac{4\rho Q}{3A}\left(\frac{2}{A}\frac{\partial Q}{\partial z}-\frac{Q}{A^{2}}\frac{\partial A}{\partial P}\frac{\partial P}{\partial z}\right).
\label{NSQPrel}
\end{align}

We now begin to eliminate terms from Eq. \ref{NSQPrel}. First, we restrict ourselves to the regime in which deviations of $A$ from its mean value of $A_{0}$ are small and linearly related to $P$. This can be expressed as $A(z,t)\approx A_{0}+cP(z,t)$ and $A_{0}\gg cP(z,t)$, where $c$ is a constant representing the compliance per unit length of the vessel. With this, we can define two parameter sets. The first are the physical parameters $r$, $\ell$, and $c$, where $r$ is the resistance per unit length and $\ell$ is the interia, while the second are the characteristic parameters $\lambda$, $\tau$, and $\alpha$. These can be defined via

\begin{subequations}
\begin{equation}
r = \frac{8\pi\mu}{A_{0}^{2}},
\label{rrel}
\end{equation}
\begin{equation}
\ell = \frac{\rho}{A_{0}},
\label{lrel}
\end{equation}
\begin{equation}
c = \frac{\partial A}{\partial P},
\label{crel}
\end{equation}
\label{rlcrels}
\end{subequations}

\begin{subequations}
\begin{equation}
\lambda = \frac{2}{r}\sqrt{\frac{\ell}{c}},
\label{lamrel}
\end{equation}
\begin{equation}
\tau = \frac{2\ell}{r}
\label{taurel}
\end{equation}
\begin{equation}
\alpha = \sqrt{\frac{c}{\ell}}.
\label{alfrel}
\end{equation}
\label{ltarels}
\end{subequations}

\noindent Of note is that while $r$ and $\ell$ have clear power law dependencies on $A_{0}$, it is not obvious how $c$ relates to $A_{0}$. However, as noted in Sec. \ref{sec:disc}, $c$ can be expressed as linearly proportional to $A_{0}$ via $c=DA_{0}$ under the already imposed condition that $A_{0}\gg cP(z,t)$. This in turn causes $\lambda$, $\tau$, and $\alpha$ to all be linearly proportional to $A_{0}$ as well. While this particular scaling is not relevant to the derivation of Eqs. \ref{Qeq}-\ref{Psym}, it does enable us to appropriately scale $\lambda$, $\tau$, and $\alpha$ between vessels of different size in the networks considered in Sec. \ref{sec:networks}.

Returning the Eq. \ref{NSQPrel}, we can use the physical parameters and move all terms to the left hand side to transform Eq. \ref{NSQPrel} into

\begin{align}
&0 = \frac{\partial P}{\partial z}\left(1-\frac{4}{3}c\ell\frac{A_{0}}{A}\left(\frac{Q}{A}\right)^{2}\right)+\ell\frac{A_{0}}{A}\frac{\partial Q}{\partial t} \nonumber\\
&+rQ\left(\frac{A_{0}}{A}\right)^{2}\left(1-\frac{1}{8\pi}\left(\frac{Q}{A}\right)^{-1}\frac{\partial^{2}Q}{\partial z^{2}}+\frac{8\ell}{3A_{0}r}\frac{\partial Q}{\partial z}\right).
\label{NSQPrlc}
\end{align}

\noindent The condition $A_{0}\gg cP(z,t)$ allows for the approximation $A_{0}/A(z,t)\approx 1-cP(z,t)/A_{0}$. Expanding each instance of $A_{0}/A(z,t)$ in this way and keeping only the constant term allows us to effectively ignore these factors in Eq. \ref{NSQPrlc}. For the pressure term, $\partial P/\partial z$, we see that there exists another source of nonlinearity. Firstly, we note that $Q/A$ represents the area averaged value of the axial velocity, which from Eq. \ref{HPQu} can be expressed as $U/2$, while as noted of Eq. \ref{Qgensol} from the main text, the velocity $1/\sqrt{c\ell}=\lambda/\tau$ is the boundary condition propagation velocity. Thus, by restricting our system to the slow regime in which the flow velocity is much slower than the pulse propagation velocity, $U/2\ll\lambda/\tau$, this nonlinear term becomes negligible in comparison to the unit term preceding it.

For the resistance term, $rQ$, there are three distinct terms in the parenthesized factor; a unit term, a linear second derivative term, and a nonlinear first derivative term. In looking at the nonlinear term first, we can expand $r$ and $\ell$ back into their constituent factors and $A_{0}$ into $\pi a_{0}^{2}$ to rewrite this term as

\begin{equation}
\frac{8\ell}{3A_{0}r}\frac{\partial Q}{\partial z} = \frac{1}{6}\left(\frac{2a_{0}\rho Q}{\mu A_{0}}\right)\left(\frac{a_{0}}{Q}\frac{\partial Q}{\partial z}\right).
\label{Rexsep}
\end{equation}

\noindent The factor of $2a_{0}\rho Q/(\mu A_{0})$, which can also be expressed as $8Q\tau/(\pi a_{0}^{3})$, is precisely the Reynolds number of the vessel. The additional factor of $(a_{0}/Q)(\partial Q/\partial z)$ represents the ratio between the vessel radius and the effective length scale over which significant changes in $Q$ occur. The combination of these factors allows this term to be made negligible in comparison to the unit term either by restricting the system to small Reynolds numbers or containing the dynamics of $Q$ such that $Q$ varies over length scales much longer than the vessel radius.

As will be shown in Appendix \ref{FSS}, when $Q$ is Fourier transformed in time, each Fourier mode can be broken into two terms which satisfy $\partial\tilde{Q}/\partial z=\pm (k(\omega\tau)/\lambda)\tilde{Q}$, where $k(\omega\tau)$ is defined in the main text as well as Eq. \ref{kdef}. Thus, the factor of $(a_{0}/Q)(\partial Q/\partial z)$ can be made negligibly small so long as $\abs{k(\omega\tau)}\ll\lambda/a_{0}$ for all frequencies that significantly contribute to $Q$. This condition not only allows Reynolds number to have a moderate magnitude but also makes the linear second derivative term negligible as well. Again invoking the Fourier space solutions gives $(\tilde{Q}/A)^{-1}(\partial^{2}\tilde{Q}/\partial z^{2})=\pi(ak(\omega\tau)/\lambda)^{2}$, which is also negligibly small when $\abs{k(\omega\tau)}\ll\lambda/a_{0}$. Yet another benefit of this restriction can be found by considering the Womersely number, $a_{0}\sqrt{\omega\rho/\mu}=2\sqrt{\omega\tau}\le\sqrt{2}\abs{k(\omega\tau)}$. Thus, so long as the dominant frequencies are sufficiently small, $\abs{k(\omega\tau)}\ll 1$, and the length scale of the system is on the order of or large than the vessel radius, $\lambda/a_{0}\gtrsim 1$, then the Womersely number will also be small, validating Eq. \ref{HPuz} as an approximation of $u_{z}$. Additionally, the unit term within the resistance term of Eq. \ref{NSQPrlc} will dominate over the other two, allowing them both to be assumed negligible.

Neglecting each of these terms of Eq. \ref{NSQPrlc} and applying the physical parameters $r$, $\ell$, and $c$ allows us to express Eqs. \ref{QArel} and \ref{NSQPrlc} as

\begin{equation}
\frac{\partial Q}{\partial z}+c\frac{\partial P}{\partial t} = 0,
\label{incomrlc}
\end{equation}

\begin{equation}
\frac{\partial P}{\partial z}+\ell\frac{\partial Q}{\partial t}+rQ = 0.
\label{NSeqrlc}
\end{equation}

\noindent Eqs. \ref{Qeq} and \ref{Peq} can be obtained from Eqs. \ref{incomrlc} and \ref{NSeqrlc} by simply expanding $r$, $\ell$, and $c$ into their constituent factors via Eq. \ref{rlcrels}. Alternatively, multiplying Eq. \ref{incomrlc} by $\lambda$ and Eq. \ref{NSeqrlc} by $\alpha\lambda$ and converting the physical parameters to the characteristic parameters then gives Eq. \ref{Qsym} and \ref{Psym} from the main text.

\section{Fourier Space Solutions}
\label{FSS}

We now denote $\tilde{Q}$ and $\tilde{P}$ as the inverse Fourier transform of $Q$ and $P$ with respect to time. This provides the relations

\begin{subequations}
\begin{equation}
\tilde{Q}\left(z,\omega\right) = \int\frac{dt}{2\pi}e^{-i\omega t}Q\left(z,t\right),
\label{QBFT}
\end{equation}
\begin{equation}
Q\left(z,t\right) = \int d\omega\> e^{i\omega t}\tilde{Q}\left(z,\omega\right),
\label{QFFT}
\end{equation}
\label{QFT}
\end{subequations}

\begin{subequations}
\begin{equation}
\tilde{P}\left(z,\omega\right) = \int\frac{dt}{2\pi}e^{-i\omega t}P\left(z,t\right),
\label{PBFT}
\end{equation}
\begin{equation}
P\left(z,t\right) = \int d\omega\> e^{i\omega t}\tilde{P}\left(z,\omega\right).
\label{PFFT}
\end{equation}
\label{PFT}
\end{subequations}

\noindent Substituting Eqs. \ref{QFFT} and \ref{PFFT} into Eqs. \ref{Qsym} and \ref{Psym} then performing the inverse Fourier transform operation on each equation thus yields

\begin{equation}
\lambda\frac{\partial\tilde{Q}}{\partial z}+i\omega\tau\alpha\tilde{P} = 0,
\label{FTdiffQeq}
\end{equation}

\begin{equation}
\lambda\frac{\partial}{\partial z}\left(\alpha\tilde{P}\right)+i\omega\tau\tilde{Q}+2\tilde{Q} = 0.
\label{FTdiffPeq}
\end{equation}

To obtain solutions for $\tilde{Q}$ and $\tilde{P}$, we first solve Eq. \ref{FTdiffQeq} for $\alpha\tilde{P}$ and substitute that into Eq. \ref{FTdiffPeq} to produce

\begin{equation}
-\frac{\lambda^{2}}{i\omega\tau}\frac{\partial^{2}\tilde{Q}}{\partial z^{2}}+\left(2+i\omega\tau\right)\tilde{Q} = 0.
\label{FTQeq}
\end{equation}

\noindent For a vessel of length $L$, Eq. \ref{FTQeq} has the general solution

\begin{equation}
\tilde{Q}\left(z,\omega\right) = \tilde{Q}_{F}\left(\omega\right)e^{-\frac{z}{\lambda}k\left(\omega\tau\right)}-\tilde{Q}_{B}\left(\omega\right)e^{-\frac{L-z}{\lambda}k\left(\omega\tau\right)},
\label{FTQgen}
\end{equation}

\noindent where

\begin{equation}
k\left(\omega\tau\right) = \sqrt{i\omega\tau\left(2+i\omega\tau\right)},
\label{kdef}
\end{equation}

\noindent and

\begin{subequations}
\begin{equation}
\tilde{Q}_{F}\left(\omega\right) = \frac{\tilde{Q}\left(0,\omega\right)e^{\frac{L}{\lambda}k\left(\omega\tau\right)}-\tilde{Q}\left(L,\omega\right)}{e^{\frac{L}{\lambda}k\left(\omega\tau\right)}-e^{-\frac{L}{\lambda}k\left(\omega\tau\right)}},
\label{FTQF}
\end{equation}
\begin{equation}
\tilde{Q}_{B}\left(\omega\right) = -\frac{\tilde{Q}\left(L,\omega\right)e^{\frac{L}{\lambda}k\left(\omega\tau\right)}-\tilde{Q}\left(0,\omega\right)}{e^{\frac{L}{\lambda}k\left(\omega\tau\right)}-e^{-\frac{L}{\lambda}k\left(\omega\tau\right)}},
\label{FTQB}
\end{equation}
\label{FTQFB}
\end{subequations}

\noindent are the forward and backward propagating current wave amplitudes. Substituting Eq. \ref{FTQgen} back into Eq. \ref{FTdiffQeq} and solving for $\tilde{P}$ then yields

\begin{align}
&\tilde{P}\left(z,\omega\right) = \nonumber\\
&\frac{k\left(\omega\tau\right)}{i\omega\tau\alpha}\left(\tilde{Q}_{F}\left(\omega\right)e^{-\frac{z}{\lambda}k\left(\omega\tau\right)}+\tilde{Q}_{B}\left(\omega\right)e^{-\frac{L-z}{\lambda}k\left(\omega\tau\right)}\right).
\label{FTPgen}
\end{align}

Eqs. \ref{FTQgen} and \ref{FTPgen} can be fully solved once sufficient boundary conditions are given. In the case where the current boundary conditions are known, $\tilde{Q}_{F}(\omega)$ and $\tilde{Q}_{B}(\omega)$ can be calculated directly from Eq. \ref{FTQFB}, thus allowing $\tilde{Q}(z,\omega)$ and $\tilde{P}(z,\omega)$ to be calculated from Eqs. \ref{FTQgen} and \ref{FTPgen}. Alternatively, when the pressure boundary conditions are known, a similar process yields the relations

\begin{align}
&\tilde{Q}\left(z,\omega\right) = \nonumber\\
&\frac{i\omega\tau\alpha}{k\left(\omega\tau\right)}\left(\tilde{P}_{F}\left(\omega\right)e^{-\frac{z}{\lambda}k\left(\omega\tau\right)}-\tilde{P}_{B}\left(\omega\right)e^{-\frac{L-z}{\lambda}k\left(\omega\tau\right)}\right),
\label{FTQgen_Pbound}
\end{align}

\begin{equation}
\tilde{P}\left(z,\omega\right) = \tilde{P}_{F}\left(\omega\right)e^{-\frac{z}{\lambda}k\left(\omega\tau\right)}+\tilde{P}_{B}\left(\omega\right)e^{-\frac{L-z}{\lambda}k\left(\omega\tau\right)},
\label{FTPgen_Pbound}
\end{equation}

\noindent where

\begin{subequations}
\begin{equation}
\tilde{P}_{F}\left(\omega\right) = \frac{\tilde{P}\left(0,\omega\right)e^{\frac{L}{\lambda}k\left(\omega\tau\right)}-\tilde{P}\left(L,\omega\right)}{e^{\frac{L}{\lambda}k\left(\omega\tau\right)}-e^{-\frac{L}{\lambda}k\left(\omega\tau\right)}},
\label{FTPF}
\end{equation}
\begin{equation}
\tilde{P}_{B}\left(\omega\right) = \frac{\tilde{P}\left(L,\omega\right)e^{\frac{L}{\lambda}k\left(\omega\tau\right)}-\tilde{P}\left(0,\omega\right)}{e^{\frac{L}{\lambda}k\left(\omega\tau\right)}-e^{-\frac{L}{\lambda}k\left(\omega\tau\right)}}.
\label{FTPB}
\end{equation}
\label{FTPFB}
\end{subequations}

Of important note is that Eqs. \ref{FTQgen}-\ref{FTPFB} are defined assuming the forward and backward waves move and decay in the forward and backward $z$ direction respectively. This forces the choice of which root to use for evaluating $k(\omega\tau)$ in Eq. \ref{kdef} to be the principle root for all real $\omega$. Eqs. \ref{FTQgen} and \ref{FTPgen} can be equivalently expressed in a way that is even in $k(\omega\tau)$ and thus independent of which root is taken. These take the forms

\begin{align}
    &\tilde{Q}\left(z,\omega\right) = \nonumber\\
    &\frac{\tilde{Q}\left(0,\omega\right)\sinh\left(\frac{L-z}{\lambda}k\left(\omega\tau\right)\right)+\tilde{Q}\left(L,\omega\right)\sinh\left(\frac{z}{\lambda}k\left(\omega\tau\right)\right)}{\sinh\left(\frac{L}{\lambda}k\left(\omega\tau\right)\right)},
    \label{FTQsol_Qbound}
\end{align}

\begin{align}
&\tilde{P}\left(z,\omega\right) = \frac{k\left(\omega\tau\right)}{i\omega\tau\alpha}\nonumber\\
&\cdot\frac{\tilde{Q}\left(0,\omega\right)\cosh\left(\frac{L-z}{\lambda}k\left(\omega\tau\right)\right)-\tilde{Q}\left(L,\omega\right)\cosh\left(\frac{z}{\lambda}k\left(\omega\tau\right)\right)}{\sinh\left(\frac{L}{\lambda}k\left(\omega\tau\right)\right)}.
\label{FTPsol_Qbound}
\end{align}

\noindent When the current boundary conditions are symmetric ($\tilde{Q}(0,\omega)=\tilde{Q}(L,\omega)$), Eq. \ref{FTQsol_Qbound} also shows that the amplitude of current oscillations at position $z$ and frequency $\omega$ obeys the steady state term of Eq. \ref{Qsymsol}.

In the limit $\omega\to 0$, Eq. \ref{FTPsol_Qbound} diverges unless $\tilde{Q}(0,0)=\tilde{Q}(L,0)$, which is guaranteed since the $\omega\to 0$ limit of Eq. \ref{FTdiffQeq} forces $\tilde{Q}(z,0)$ to be invariant with respect to changes in $z$. However, in cases where the limit of $\tilde{Q}(z,\omega)$ as $\omega\to 0$ is not well defined, such as when $Q(z,t)$ is a pulsatile function with discrete frequencies, Eq. \ref{FTPsol_Qbound} becomes equally ill-defined. This is a consequence of the fact that pressure is a gauge variable and globally changing the pressure across the whole vessel and at all times only affects $\tilde{P}(z,0)$ while having no impact on the current. Thus, while Eq. \ref{FTPsol_Qbound} retains a linear dependence on $z$ from the hyperbolic cosine terms, the $z$-independent constant term must be determined by choice of gauge. This issue can be circumvented by defining the pressure boundary conditions instead, as the gauge choice would be included into the boundary conditions. This allows for the current and pressure Fourier transforms to be expressed as

\begin{align}
&\tilde{Q}\left(z,\omega\right) = \frac{i\omega\tau\alpha}{k\left(\omega\tau\right)}\nonumber\\
&\cdot\frac{\tilde{P}\left(0,\omega\right)\cosh\left(\frac{L-z}{\lambda}k\left(\omega\tau\right)\right)-\tilde{P}\left(L,\omega\right)\cosh\left(\frac{z}{\lambda}k\left(\omega\tau\right)\right)}{\sinh\left(\frac{L}{\lambda}k\left(\omega\tau\right)\right)},
\label{FTQsol_Pbound}
\end{align}

\begin{align}
&\tilde{P}\left(z,\omega\right) = \nonumber\\
&\frac{\tilde{P}\left(0,\omega\right)\sinh\left(\frac{L-z}{\lambda}k\left(\omega\tau\right)\right)+\tilde{P}\left(L,\omega\right)\sinh\left(\frac{z}{\lambda}k\left(\omega\tau\right)\right)}{\sinh\left(\frac{L}{\lambda}k\left(\omega\tau\right)\right)}.
\label{FTPsol_Pbound}
\end{align}

\noindent Unlike Eq. \ref{FTPsol_Qbound}, Eq. \ref{FTQsol_Pbound} is well defined in the limit $\omega\to 0$ regardless of the behaviour of $\tilde{P}(0,\omega)$ and $\tilde{P}(L,\omega)$.

\section{Network Laplacian}
\label{NL}

For a network of vessels that obey the connectivity laws given by Eqs. \ref{Pconnect} and \ref{Qconnect}, we can define the pressure at node $\mu$ as $P_{\mu}(t)=P_{\mu\nu}(0,t)=P_{\nu\mu}(L_{\mu\nu},t)$ for $\nu\in\mathcal{N}_{\mu}$. This notation allows Eq. \ref{Qconnect} to be Fourier transformed and Eq. \ref{FTQsol_Pbound} to be substituted in to produce

\begin{align}
&\tilde{H}_{\mu}\left(\omega\right) = \sum_{\nu\in\mathcal{N}_{\mu}}\tilde{Q}_{\mu\nu}\left(0,\omega\right) \nonumber\\
&= \sum_{\nu\in\mathcal{N}_{\mu}}\frac{i\omega\tau_{\mu\nu}\alpha_{\mu\nu}}{k\left(\omega\tau_{\mu\nu}\right)}\frac{\tilde{P}_{\mu}\left(\omega\right)\cosh\left(\frac{L_{\mu\nu}}{\lambda_{\mu\nu}}k\left(\omega\tau_{\mu\nu}\right)\right)-\tilde{P}_{\nu}\left(\omega\right)}{\sinh\left(\frac{L_{\mu\nu}}{\lambda_{\mu\nu}}k\left(\omega\tau_{\mu\nu}\right)\right)} \nonumber\\
&= \sum_{\nu\in\mathcal{N}_{\mu}}\mathcal{L}_{\mu\nu}\left(\omega\right)\tilde{P}_{\nu}\left(\omega\right),
\label{FTHPrel}
\end{align}

\noindent where

\begin{align}
\mathcal{L}_{\mu\nu}\left(\omega\right) = &\delta_{\mu,\nu}\left(\sum_{\xi\in\mathcal{N}_{\mu}}\frac{i\omega\tau_{\mu\xi}\alpha_{\mu\xi}\cosh\left(\frac{L_{\mu\xi}}{\lambda_{\mu\xi}}k\left(\omega\tau_{\mu\xi}\right)\right)}{k\left(\omega\tau_{\mu\xi}\right)\sinh\left(\frac{L_{\mu\xi}}{\lambda_{\mu\xi}}k\left(\omega\tau_{\mu\xi}\right)\right)}\right) \nonumber\\
&-\frac{i\omega\tau_{\mu\nu}\alpha_{\mu\nu}}{k\left(\omega\tau_{\mu\nu}\right)\sinh\left(\frac{L_{\mu\nu}}{\lambda_{\mu\nu}}k\left(\omega\tau_{\mu\nu}\right)\right)}.
\label{Lmndef}
\end{align}

For nonzero $\omega$, $\mathcal{L}_{\mu\nu}$ defines an invertible matrix, meaning that for any set of input $\tilde{H}_{\mu}$ there exists a unique set of potentials $\tilde{P}_{\mu}$ that solves Eq. \ref{FTHPrel} and vice versa. Thus, there is no restriction on the oscillatory components of $H_{\mu}(t)$. However, in the $\omega\to 0$ limit, Eq. \ref{Lmndef} reduces to

\begin{align}
&\lim_{\omega\to 0}\mathcal{L}_{\mu\nu}\left(\omega\right)  \nonumber\\
&= \delta_{\mu,\nu}\left(\sum_{\xi\in\mathcal{N}_{\mu}}\frac{\alpha_{\mu\xi}\lambda_{\mu\xi}}{L_{\mu\xi}}\cdot\lim_{\omega\to 0}\left(\frac{i\omega\tau_{\mu\xi}}{\left(k\left(\omega\tau_{\mu\xi}\right)\right)^{2}}\right)\right) \nonumber\\
&\quad -\frac{\alpha_{\mu\nu}\lambda_{\mu\nu}}{L_{\mu\nu}}\cdot\lim_{\omega\to 0}\left(\frac{i\omega\tau_{\mu\nu}}{\left(k\left(\omega\tau_{\mu\nu}\right)\right)^{2}}\right) \nonumber\\
&= \delta_{\mu,\nu}\left(\sum_{\xi\in\mathcal{N}_{\mu}}\frac{\alpha_{\mu\xi}\lambda_{\mu\xi}}{2L_{\mu\xi}}\right)-\frac{\alpha_{\mu\nu}\lambda_{\mu\nu}}{2L_{\mu\nu}} \nonumber\\
&= \delta_{\mu,\nu}\left(\sum_{\xi\in\mathcal{N}_{\mu}}\frac{1}{r_{\mu\xi}L_{\mu\xi}}\right)-\frac{1}{r_{\mu\nu}L_{\mu\nu}}.
\label{Lmnlim}
\end{align}

\noindent $\mathcal{L}_{\mu\nu}$ as defined in Eq. \ref{Lmnlim} gives a noninvertible matrix which describes a transformation that takes the vector space $\mathbb{R}^{N}$, where $N$ is the number of nodes in the network, to the subspace $\mathbb{S}\subset\mathbb{R}^{N}$ defined such that $\mathbb{S}$ is the space of all vectors $v\in\mathbb{R}^{N}$ whose components sum to 0. Thus, the $\omega\to 0$ limit of Eq. \ref{FTHPrel} can only be solved if the components of $\tilde{H}_{\mu}(0)$ sum to 0. This is equivalent to the restriction that the constant components of $H_{\mu}(t)$ must all sum to 0 so that on average the current going into the network equals the current coming out.

\end{document}